\documentclass[12pt,a4paper]{article}
\usepackage{jheppub}

\usepackage{epsfig,multirow,subfigure} 
\usepackage{amscd}
\usepackage[matrix,arrow,curve]{xy}
\usepackage{verbatim}
\usepackage{latexsym}
\usepackage{amsfonts,amsthm,amsmath,mathtools}
\newcommand{\be}{\begin{equation}}
\newcommand{\ee}{\end{equation}}
\newcommand{\bea}{\begin{eqnarray}}
\newcommand{\eea}{\end{eqnarray}}

\newcommand{\Tr}{{\rm Tr}}

\renewcommand{\l}{\lambda}

\newcommand{\dd}{{\rm d}}

\subheader{UCSD-PTH-11-19}
\title{\begin{center}
The Large N Limit of Toric Chern-Simons Matter Theories  and Their Duals
\end{center}}
\author[a]{Antonio Amariti,}
\author[b]{Claudius Klare,}
\author[c]{Massimo Siani}

\affiliation[a]{Department of Physics, University of California, \\
San Diego La Jolla, CA 92093-0354, USA}
\affiliation[b]{Dipartimento di Fisica, Universit\`a di Milano-Bicocca, I-20126 Milano, Italy \\
                and \\
	       INFN, sezione di Milano-Bicocca, I-20126 Milano, Italy}
\affiliation[c]{Instituut voor Theoretische Fysica, Katholieke Universiteit Leuven,\\
Celestijnenlaan 200D, B-3001 Leuven, Belgium \\ and \\
Department of Particle Physics and Astrophysics \\
Weizmann Institute of Science, Rehovot 76100, Israel}

\emailAdd{amariti@physics.ucsd.edu}
\emailAdd{claudius.klare@mib.infn.it}
\emailAdd{massimo@weizmann.ac.il}

\abstract{We compute the large $N$ limit of the localized three dimensional free energy
of various field theories with known proposed AdS duals. We show that vector-like theories agree with the expected supergravity
results, and with the conjectured $F$-theorem.  We also check that the
large N free energy is preserved by the three dimensional Seiberg duality for general classes of vector like theories. 
Then we analyze the behavior of the free energy 
of chiral-like theories by applying a new proposal. The proposal is based 
on the restoration of a discrete symmetry on the free energy before the extremization.
We apply this procedure at
strong coupling in some examples and we discuss the results.
We conclude the paper by proposing an alternative  geometrical expression for the free energy.}

\begin{document}

\maketitle

\section{Introduction} \label{intro}

Recently, three dimensional field theories on curved backgrounds gained new attraction from the observation that the partition function
localized  on  $S^3$
can be reduced to a matrix integral, providing an \emph{exact} quantity of the quantum theory.
This was established in \cite{Kapustin:2009kz}
by generalizing the four dimensional results of \cite{Pestun:2007rz}, 
and allowed  many  checks of known and expected
results of three dimensional field theories.

One of the most attractive consequences is the possibility to compute in field theory some 
results already known  from the gravity dual in the  AdS$_4/$CFT$_3$ correspondence
\cite{Bagger:2006sk,Bagger:2007jr,Bagger:2007vi,Gustavsson:2007vu,Aharony:2008ug}.
This should provide a non trivial test of the correspondence itself.

The AdS$_4$/CFT$_3$ correspondence relates $M$-theory on AdS$_4 \times Y$ to a three dimensional SCFT
describing the IR dynamics of M$2$ branes probing a Calabi-Yau cone $X=C(Y)$ over the 
seven dimensional Sasaki-Einstein manifold $Y$.
 
In \cite{Bagger:2006sk,Bagger:2007jr,Bagger:2007vi,Gustavsson:2007vu} 
it was shown that a Chern-Simons (CS) matter theory is necessary to describe the low energy theory on $M2$ branes, 
which eventually lead to the ABJM theory \cite{Aharony:2008ug}  for $N$ $M2$ branes in $X=\mathbb{C}^4/\mathbb{Z}_k$.
Cases with lower supersymmetry can be studied by modifying the manifold $X$.

The localized partition function on $S^3$ gives us a further handle to check this construction.
As a prominent example,  in \cite{Drukker:2010nc} the predicted $N^{3/2}$ free energy scaling of gravity backgrounds 
generated by M$2$-branes has been  reproduced in the strongly coupled field theory side of the ABJM  model.
Furthermore, in \cite{Herzog:2010hf} the authors observed a direct relation between the volume of the
base of the Calabi-Yau space $X$ transverse to a stack of $N$ M$2$ branes
and the large $N$ partition function of the corresponding $\mathcal{N}=3$ field theory on the branes world volume.

In the $\mathcal{N}=2$ case the  mass dimensions of the superfields  at the infrared fixed point
are not fixed by supersymmetry, due to the mixing of the $R$ charge with the abelian symmetries of the theory.
The
partition function for an arbitrary choice of R-charges in $\mathcal{N}=2$ theories
was first computed in   \cite{Jafferis:2010un,Hama:2010av}.
In 
\cite{Jafferis:2010un} it was shown that the partition function $\mathcal{Z}$, 
 computed on a three sphere, is extremized by the \emph{exact} superconformal $R$-symmetry.
Then it was observed that the free energy $F=-\log |\mathcal{Z}|$ 
has a monotonic decreasing  behavior along the  the RG flow, and this led to conjecture 
the existence of an $F$-theorem  \cite{Jafferis:2011zi,Klebanov:2011gs}.

This technique has been
successfully applied to a wide spectrum of three dimensional field theories \cite{Martelli:2011qj,Cheon:2011vi,Jafferis:2011zi,Amariti:2011hw,Niarchos:2011sn,Minwalla,Amariti:2011da,
Morita:2011cs,Benini:2011cma,Amariti:2011xp}, both at weak and at strong coupling.
The former computations can be matched with the standard diagrammatic evaluations. The latter provide a way to
test the proposed AdS/CFT dual pairs. The supergravity dual predicts that the free energy 
at leading order in a large $N$ expansion is given by $F\simeq N^{3/2}$ Vol$(Y)^{-1/2}$.
Thus,
one may compute the free energy from the field theory side
and check the proposed correspondence by comparison with the volume of the transverse geometry.
A large class of AdS/CFT dual pairs has passed this nontrivial test
\cite{Martelli:2011qj,Cheon:2011vi,Jafferis:2011zi}. However,
every field theory considered so far contains an equal number of bifundamental and anti-bifundamental fields for each
gauge group. By borrowing the four-dimensional language, they are called vector-like (or non-chiral) theories.
Another class of theories contains a different number of bifundamental and anti-bifundametal fields for some of
the gauge groups (chiral like gauge theories). Previous results in these cases showed that the same techniques used in the non-chiral computations
do not lead to the scaling $F \propto N^{3/2}$, but instead to $F \propto N^2$. Understanding whether this is only an artifact
of the applied techniques is one of the aims of this paper.

In this note we discuss  several aspect of the AdS$_4/$CFT$_3$ correspondence for  
$\mathcal{N}=2$ SCFTs with the help of the localized free energy.

We start by discussing several vector-like theories.
It was observed in many examples \cite{Martelli:2011qj,Jafferis:2011zi,Cheon:2011vi,Gulotta:2011aa}
that the free energy with arbitrary $R$-charges and the volumes as functions of the geometrical data
are the same even before extremization.
Here we generalize this result and show that 
it is natural to consider the  meson generating function,
which intrinsically encodes the information on the global symmetries and reproduces the free energy
as a function of the charges under these symmetries.
This is an analogous observation to what has been proven  in four dimensions \cite{Butti:2005vn,Eager:2010yu}.
We observe further, that the free energy decreases along an RG flow connecting the theories that we consider,
corroborating the validity of the conjectured $F$-theorem.
On the field theory side, the RG flow corresponds to giving an expectation value
to one of the scalar fields and then integrating it out.
This has a counterpart on the gravity side related to partial resolution of the singularity.

Then we discuss Seiberg duality in vector like theories with multiple gauge groups.
We observe that the large $N$ free energy is preserved even before extremization, as in the 
$\mathcal{N}=3$ case \cite{Herzog:2010hf} under the rules derived in \cite{Aharony:2008gk,Giveon:2008zn,Amariti:2009rb}.

We then switch to the analysis of large $N$ chiral like quiver gauge theories,
and its relation with the volumes of the proposed dual toric Sasaki-Einstein manifolds.
We apply the technique recently discussed in \cite{Amariti:2011jp}, where it was observed that a consistent large $N$ limit
in the chiral-like models needs the free energy to be rewritten in a manifestly symmetric form. Indeed, there exists a
hidden reflection symmetry acting on the Cartan sub-algebra of the gauge group whose manifest appearance is crucial, in the
large $N$ limit, to reproduce the leading order behavior of the free energy itself.
One can then think to extend this conjecture in the strongly coupled regime, and see if the expected large $N$ scaling 
properties are recovered even in that limit.
Even though such an extension is non-trivial and we leave some open questions for further study,
we observe  that there are at least two models described by chiral like quiver gauge theories
which fit with this procedure, namely $M^{111}/\mathbb{Z}_k$  and 
$Q^{222}/\mathbb{Z}_k$. 
Many quiver gauge theories have been conjectured to describe the motion of M$2$ branes probing these singularities \cite{
Franco:2008um,Aganagic:2009zk,Franco:2009sp,Hanany:2008fj,Amariti:2009rb,Davey:2009sr}.
Here we concentrate to the phases that share the same field content and superpotential of
 a stack of D$3$ branes probing  $\mathbb{C}^3/\mathbb{Z}_3$ and 
 $\mathbb{F}_{0}^{(I)}$ respectively.
We show that once the monopole charge is set to zero the value of the extremized free energy
corresponds to the AdS/CFT volumes computation.

The last topic that we discuss is related to the construction of a pure field theoretical quantity from the geometrical data, along the lines of 
\cite{Butti:2005vn}. Indeed in that paper it was shown that the central charge $a$ can be obtained directly from the information of the dual geometry.
We show that the generalization does not follow straightforwardly. By exploiting the symmetries of toric Calabi-Yau four-folds,
we give a procedure  to generalize the cubic formula
of \cite{Butti:2005vn,Lee:2006ru} to three-dimensional field theories. We apply our general discussion to many examples and find,
quite surprisingly, a formula that is quartic in the $R$ charges and reproduces the field theory computations by only using
the geometrical data and without any reference to localization.

The paper is organized as follows. In Section \ref{sec:N2toric} we review the computation of the moduli
space of toric ${\cal N}=2$ gauge theories and review the methods to extract the geometrical data from the
field theory. In Section \ref{sec:Fvector} we compute the localized free energy in vector-like models and
compare with the geometric dual predictions.
The Seiberg dual phases of a large class of vector-like theories with multiple gauge groups are discussed in Section
\ref{sec:Seiberg} and the duality from the large $N$ free energy point of view is presented.
In Section \ref{sec:Chirals} we explain our approach to
the large $N$ extremization of the free energy in chiral-like models and apply it to some examples.
We comment on a different formulation of the extremization problem in field theory and on its relation to
the volume minimization in Section \ref{sec:Rquartic}.
We conclude by discussing our results and by outlining possible directions for further research.

\vspace{.5cm}
{\bf Note added:} section 3 of \cite{Gulotta:2011vp}, which appeared on the same day as this paper, significantly overlaps with our
section \ref{sec:Seiberg}.

\section{$\mathcal{N}=2$ CS toric quivers and their moduli space}
\label{sec:N2toric}

\subsection{The field theory description}
\label{sec:N2FieldTheory}
In this section we briefly review the main aspects of the gauge theories
that we study in the rest of the paper.
They are three dimensional ${\cal N}=2$ supersymmetric quiver gauge theories which are believed to describe
the low energy dynamics of a stack of M$2$ branes probing a toric CY$_4$ singularity.
We consider a product  gauge group $\prod U(N_a)$ such that the corresponding gauge fields have 
a Chern-Simons term at level $k_a$.
We add matter fields either in the bifundamental or in
the adjoint representation of the gauge groups. In the ${\cal N}=2$ language, the Lagrangian reads
\bea \label{Lag}
\mathcal{L} &=& 
\sum_a \frac{k_a}{2\pi} \int d^4 \theta \int_0^1 dt \, V_a \overline D^\alpha (e^{-t V_a} D_\alpha e^{t V_a})+
\int d^4 \theta \, \sum_{X_{ab}} X^{\dagger}_{ab} e^{V_a} X_{ab} e^{-V_b} \nonumber \\
&+& \int d^2 \theta \, W(X)+c.c.
\eea
The first term is the CS Lagrangian at level $k_a$ for the gauge superfield $V_a$ associated to the gauge
group $U(N_a)_{k_a}$. The second term is the usual minimal coupling between matter and gauge fields,
and $W$ is the superpotential. We will be interested in toric field theories, where every matter field
appears exactly twice in the superpotential,
once in a term with a positive sign and once in a term with a negative sign. This is the toric condition, which
highly constrains the space of solutions to the F-terms.
In three dimensions, in the WZ gauge, the vector superfield is  
\be
V = i \theta \bar \theta \, \sigma + \theta \gamma^{\mu} \bar \theta \, A_{\mu} - \theta^2 \bar\theta \, \bar\l
- \bar \theta^2 \theta \, \l + \theta^2 \bar \theta^2 \, D
\ee
where we drop fermionic indices. The field 
 $\sigma$ is an auxiliary scalar coming from the dimensional reduction  four dimensional gauge field.
In terms of the component fields, the classical Chern-Simons Lagrangian becomes
\be
S_{CS} = \sum_a \frac{k_a}{4 \pi} \int \Tr \left(
A_a \wedge \text{d} A_a+\frac{2}{3} A_a \wedge A_a \wedge A_a 
-\overline \l_a \l_a + 2 D_a \sigma_a 
\right)
\ee
The classical moduli space  for unbroken supersymmetry
is obtained by minimizing the scalar potential, which is equivalent to the vanishing of the
F- and D-terms
\be \label{EOM}
\begin{split}
\partial_{X_{ab}} W =0 \\
\mu_a(X) \equiv \sum_b X_{ab}X_{ab}^{\dagger} - \sum_c X_{ca}^{\dagger} X_{ca}
+[X_{aa},X_{aa}^{\dagger} ]=4 k_a \sigma_a \\
\sigma_a X_{ab} - X_{ab} \sigma_b=0
\end{split}
\ee 
In this paper we study the moduli space for the abelian case $N_a=N=1$. This is the  mesonic moduli space
$\mathcal{M}$ and corresponds in the gravity dual to the transverse space of a single M$2$-brane.
In all our examples, the latter is a four-dimensional toric Calabi-Yau cone. The moduli space of the
non-abelian theories can be obtained by taking the $N$-th symmetric product of $\cal M$ \cite{Martelli:2008si,Hanany:2008cd,Jafferis:2008qz}.
We start by solving the F-term equations given in the first line in \eqref{EOM}.
The solutions of these equations define the master space $ ~^{Irr}\mathcal{F}^{\mathbf{b}}$ \cite{Forcella:2008bb},
whose main irreducible component is a toric variety of dimension $G+2$,  where $G$ is the number of gauge groups.
The remaining three dimensional equations of motion turn out to be slightly more involved than the
four-dimensional ones, because of the scalar auxiliary field $\sigma$.
First, we see that the third equation in (\ref{EOM}) sets every $\sigma_a=\sigma$ if we want to avoid trivial
solutions.
Furthermore, since the overall gauge group decouples,
we have to choose the Chern-Simons levels such that 
\be
\sum k_a = 0 \, ,
\ee
otherwise the mesonic moduli space is three dimensional \cite{Martelli:2008si,Hanany:2008cd},
and cannot describe the transverse space to a M$2$ brane. 
This leaves only $G-1$ independent equations out of the $G$
in the second line of (\ref{EOM}).
We take $G-2$ nontrivial linear combinations of the independent moment maps $\mu_a$ such
that each linear combination vanishes using \eqref{EOM}; these identify directions in the gauge
space orthogonal to the $k$'s and they correspond to canonical D-terms. 
They are automatically solved if we impose gauge invariance under
the complexified gauge group. 
The remaining equation, which identifies the parallel direction to the $k$'s, sets the value of $\sigma$,
which does not affect the following discussion. 
Furthermore, one can show that the corresponding gauge group is broken to a discrete subgroup
and that it is not imposed as a continuos gauge symmetry.
The mesonic moduli space $\cal M$
is obtained  by modding out the irreducible component of the master space  by the $G-2$ gauge groups described above.
Formally, it can be written as
\be
\mathcal{M} = ~^{Irr}\mathcal{F}^{\mathbf{b}}/H
\label{eq:MasterModuli}
\ee
where $H$ is the $(\mathbb{C}^*)^{G-2}$ kernel of 
\be
C = \left(\begin{array}{cccccc}
1&1&1&1&1&1\\
k_1&k_2&\dots&\dots&k_{G-1}&k_G
\end{array}
\right)
\label{eq:Ck}
\ee

\subsection{The toric description}
\label{sec:TheToricDescription}

In the case of toric quiver gauge theories, the information about the moduli space of the field theory
is encoded in a set of combinatorial data which are represented through the so-called toric diagram.
For most purposes, the latter is the only object one needs in comparing field theoretical and geometrical
quantities, and it can be extracted in many ways. The algorithm we will use heavily relies on the results
presented in \cite{Franco:2008um,Hanany:2008gx}.

In order to obtain the toric diagram from the field theory data,
we construct the so-called perfect matching matrix in two steps, as follows. Due to the toric condition, there is an even number
of superpotential terms, half of them come with a positive sign, and the other half with a negative sign.
Moreover, every field $X$ appears exactly once in each set of terms, say in the $i$-th term of the positive
set and in the $j$-th term of the negative set.
We construct a matrix by adding the fields X to the $(i,j)$-th entries.
The determinant $K$ of this matrix is a polynomial with $c$ terms.
Once again we construct a matrix,
this time the $(i,j)$ entry is $1$ if the $i$-th field is in the $j$-th term of $K$, and $0$ otherwise. 
This is is the perfect matching matrix, which we denote $P$.
We can decompose the fields as $X_a = \prod_{\alpha=1}^{c} p_{\alpha}^{P_{a \alpha}}$,
where the $p_{\alpha}$'s are called perfect matchings. This decomposition automatically solves the $F$-term equations.

Then we define the incidence matrix $d$ of the quiver. Each row corresponds to a gauge group, and each
column to a field. The $(i,j)$ entry is $1$ if the $j$-th field transforms in the fundamental representation
of the $i$-th gauge group, $-1$ if the field transforms according to the antifundamental representation, and
$0$ otherwise. By using the incidence matrix and the perfect matching matrix we can define a new matrix $Q$
by
\be
d = Q \cdot P^T
\ee
It is the charge matrix of the associated GLSM \cite{Witten:1993yc}, and gives the D-terms when modded out by the gauge symmetry.
Similarly, the perfect matching matrix $P$, extracted from the superpotential alone, gives the F-terms.
Putting this together, the toric diagram for the complex four-dimensional Calabi-Yau cone is given by
\be \label{TD}
G_T = {\rm Ker}\left(
\begin{array}{c}
Q_F\\
Q_D
\end{array}
\right)^T \equiv
{\rm Ker}\left(
\begin{array}{c}
{\rm Ker}(P)^T\\
{\rm Ker}(C) \cdot Q
\end{array}
\right)^T
\ee
where $C$ is given in \eqref{eq:Ck}. $G_T$ is a matrix with four rows and the $n$ columns are the
$n$ four-vectors generating the fan for the four-dimensional toric Calabi-Yau cone.
Every column of this matrix is in one-to-one correspondence with the perfect matchings  represented as the $c$ columns
of $P$ or the $c$ terms in $K$.
By a $SL(4,\mathbb{Z})$
transformation we can rotate all the vectors such that each last component is $1$, viz.\ $v_i=(w_i,1)$, $i=1,\dots,n$.
This is due to the Calabi-Yau condition.
The convex hull of the $w$'s is the toric diagram. 

The toric diagram encodes all the data about the toric Calabi-Yau and
its base, which by definition is a toric Sasaki-Einstein space. In particular, we can compute the volume of
the base and of its five-cycles by only looking at the vectors in the matrix $G_T$. Each independent
compact five-cycle is in correspondence with an external point of the toric diagram and its volume is
a function of the Reeb vector $b$ \cite{Martelli:2005tp},
 a constant norm Killing vector field commuting with all the  isometries  $Y$.
Let $v_i=(w_i,1)$ be the vector in the toric fan, corresponding to the external point $w_i$ in the diagram 
and consider the counter-clockwise ordered sequence of vectors $w_k, k=1,\ldots , n_i$, that are adjacent to $v_i$.
We can compute 
the volume of a  $5$-cycle $\Sigma_i$
on which a M$5$ brane is wrapped as 
\cite{Martelli:2005tp,Hanany:2008fj}
\begin{equation}
\begin{split}
  \mathrm{Vol}(\Sigma_i) &= \sum_{k=2}^{n_{i}-1}  \frac{
(v_i,w_{k-1},w_k,w_{k+1})(v_i,w_k,w_1,w_{n_i})}{(v_i,b,w_k,w_{k+1})
(v_i,b,w_{k-1},w_k)(v_i,b,w_1,w_{n_i})}
\end{split}
\label{MSvol}
\end{equation}
and define the sum of these volumes as 
\begin{equation}
Z = \sum_i {\rm Vol}(\Sigma_i)
\label{MSvol2}
\end{equation}
where $(v_1,v_2,v_3,v_4)$ denotes the determinant of four vectors $v_{1,2,3,4}$. 
The value of the Reeb vector which minimizes the volume functional (\ref{MSvol2})
gives rise to the Calabi-Yau metric.
Note that imposing the Calabi-Yau condition via $v_4=1$, 
implies setting the fourth component of the Reeb vector $b_4=4$.

A five-brane wrapped on a given five-cycle $\Sigma_i$ corresponds to an operator with dimension \cite{Fabbri:1999hw}
\be \label{eq:DeltaofR} \Delta_i= \frac{2 {\rm Vol}(\Sigma_i)}{Z} \ee
In the next section we present yet another way to compute the volume functional of the
underlying moduli space geometry, directly from the field theory data but having a very natural
interpretation in the toric language.

\subsection{The Hilbert series}
\label{sec:FeqHil}

A convenient way to extract the volume of the moduli space, 
which does not require the geometrical technologies involving the Reeb vector and individual $5$-cycles,
is related to counting the mesonic operators.
The counting can be performed by the Hilbert series, 
which is the partition function for the mesons on the M$2$ moduli space,
see e.g.\ \cite{Benvenuti:2006qr,Forcella:2008bb,Hanany:2008fj}.
The pole of the series gives the demanded volume, 
while keeping track of the dependency on the global symmetries.
In the toric case the counting becomes particularly easy,
since we we can systematically solve the $F$-terms through perfect matchings,
which results in the quotient description of the moduli space
\begin{equation}
  \mathcal{M} =  \mathbb{C}^d / \left( \mathbb{C}^* \right)^{d-4}  \,.
  \label{eq:HilGLSM}
\end{equation}
Here, $d$ denotes the number of perfect matchings assigned to external points of the toric diagram
and the charge matrix of the quotient is given by the linear relations amongst the corresponding vectors in the fan.
This  quotient construction makes manifest the dependency 
on all of the global symmetries, 
which the moduli space inherits from the natural isometries of the ambient space $\mathbb{C}^{d}$.
Generically, there are more external perfect matchings than global symmetries.
This is  because we had to introduce extra fields together with spurious symmetries, which are not seen by the physical fields,
when solving the $F$-terms via perfect matchings.
Upon parameterizing the symmetries by the perfect matchings 
we might encounter a redundancy.

The Hilbert series for the flat ambient space reduces to the geometrical series $\mathrm{Hil} \sim 1/(1-t)^d$,
and the quotient can be realized by projecting on the singlets under the $\left(\mathbb{C}^\ast\right)^{d-4}$ action,
\begin{equation}
  \mathrm{Hil}(t_i; \mathcal{M}) = \oint \prod_{k=1}^{d-4} \frac{\dd z_k}{2\pi i z_k} \frac{1}{\prod_{i=1}^{d} (1-t_i Z_i)} \,,
  \label{eq:Hilbert}
\end{equation}
where $Z_i = Z_i(z_k)$ is the monomial weight of the $i$-th homogeneous coordinate
under the $\left(\mathbb{C}^\ast\right)^{d-4}$ action in $\eqref{eq:HilGLSM}$.
If we further set $t_i=e^{-2 \epsilon a_i}$ and take the $\epsilon \rightarrow 0$ limit,
we have \cite{Martelli:2006yb}
\begin{equation}
  \text{Hil}(t_i; \mathcal{M}) \sim \frac{\mathrm{Vol}(a_i; Y)}{\epsilon^4} + \dots \; ,
\end{equation}
which gives us an expression for the volume of the base $Y$ 
in terms of charges under the global symmetries, corresponding to the external perfect matchings.

\section{Vector-Like models}
\label{sec:Fvector}

\subsection{The free energy of large $N$ vector-like quivers}

In this section we discuss the computation of the leading order term of the free energy
of vector-like field theories in a large $N$ expansion.
The localized partition function on the three-sphere
reads
 \begin{equation}
 {\cal Z} = \int  \text{d}\left[ \frac{\lambda}{2\pi}\right] \prod_a \left[
e^{\frac{ i \sum k_a {\lambda_i^{(a)}}^2 }{4 \pi}- \sum \Delta_m^{(a)} \lambda_i^{(a)} } \prod_{i<j} 
  \sinh^2
 \left(\frac{\l_i^{(a)}-\l_j^{(a)}}{2}\right)
 \prod_{\rho}e^{l\left(1-\Delta+i \rho\left(\frac{\l}{2 \pi}\right)\right)} \right]
\label{eq:Zgen}
 \end{equation}
where the integral extends over the $\sum N_a$ variables $\l_i^{(a)}$, $k_a$ are the Chern-Simons levels in the Lagrangian,
$\Delta_m^{(a)}$ is the bare monopole charge associated with the $a$th gauge group and
$\ell(z)$ is the one loop determinant of the matter fields computed in \cite{Jafferis:2010un,Hama:2010av} 
\begin{equation}
\ell(z)=-z \log\left(1-e^{2 \pi  i z}\right)+\frac{i}{2} \left(\pi  z^2 + \frac{1}{\pi} Li_2 e^{2 \pi  i z}\right) - \frac{i \pi}{12}
\end{equation}
with derivative
\be
\ell^\prime(z) = - \pi z \cot\left(\pi z\right)
\label{eq:lder}
\ee
and $\rho$ refers to the
weights of the representation of every single matter field.

Following the presentation of  \cite{Jafferis:2011zi},
 we restrict to theories with a product gauge group $\prod U(N_a)_{k_a}$, 
 at large $N$ and with $\sum k_a=0$.
The integral at large $N_a$ and finite $k_a$ is dominated by the minimum of the free energy
$F=-\log{|\mathcal{Z}|}$. The equations of motion $\partial_{\l_i^{(a)}} F =0$ contain two kinds of
contributions that act on the eigenvalues \cite{Martelli:2011qj,Jafferis:2011zi}, dubbed short range
and long range forces.
The latter are defined as those contributions that can be approximated with
the sign of ${\rm Re}(z)$ in \eqref{eq:Zgen}, and cancel out in vector-like theories which satisfy $\sum k_a=0$
and where the eigenvalues are given by
\begin{equation}
\lambda_i^{(a)} = N^{1/2} x_i + i y_i^{(a)}
\label{eq:eigen}
\end{equation}
For large enough $N$, one can replace the discrete set \eqref{eq:eigen} with $|G|$ continuous variables.
The real part of the eigenvalues becomes a dense set with density
$\rho(x)=ds/dx$ and the imaginary parts $y_i^{(a)} $ become
the functions $y_a(x)$.
One finds that the free energy 
is given by two contributions, one is the classical 
one from the Chern-Simons and monopole terms
\begin{equation} \label{integral}
F_{CS} = \frac{N^{3/2}}{2 \pi} \int {\rm d}x\, \rho(x) \, x\, \sum_{a} \left( k_a y_a + 2\pi \Delta_m^{(a)} \right)
\end{equation}
while the second contribution comes from the one loop determinant of the vector and the matter fields.
The former actually vanishes and we are left with the latter.
In vector-like theories, for a pair of bifundamental and anti-bifundamental fields with dimensions
$\Delta_{ab}$ and $\Delta_{ba}$ respectively, we have
\begin{equation} \label{oneloop}
F_{1-loop} =-N^{3/2}\frac{(2-\Delta_{ab}^+)}{2} \int {\rm d}x \rho(x)^2 \left(\delta y_{ab}^2 -\frac{\pi^2}{3} \Delta_{ab}^+ (4-\Delta_{a b}^+) \right)
\end{equation}
where $\delta y_{ab} \equiv y_a(x)-y_b(x)+\pi \Delta_{ab}^-$ and $\Delta_{ab}^\pm=\Delta_{ab} \pm \Delta_{ba}$.
For an adjoint field we use (\ref{oneloop}) with $a=b$ and divide by a factor two.
Equation \eqref{oneloop} is only valid in the range $|\delta y_{ab}| \leq \pi \Delta^+_{ab} $. While
our solutions will always respect this constraints, one should note that in general the free energy is
not a differentiable function at $|\delta y_{ab}| = \pi \Delta^+_{ab} $.

The resulting free energy has to be extremized as a functional of $\rho$ and the $y$'s. The former has to satisfy the
constraints
\be
\int \dd x\, \rho(x) =1 \qquad \rho(x) \geq 0 \,\,\, {\rm pointwise}
\nonumber
\ee
to be interpreted as an eigenvalue density. We will impose the former constraint through a Lagrange multiplier $\mu$.
This set of rules is enough to compute the free energy of the 
vector like theories as a function of the $R$ charges.

As observed in \cite{Jafferis:2011zi} the  expressions (\ref{integral}) and \eqref{oneloop} possess flat directions
which parameterize the symmetries on the eigenvalues and on the $R$ charges. 
By defining the real parameters $\eta^{(a)}$  they are
\begin{eqnarray}\label{flat}
&&y_a\rightarrow y_a - 2 \pi \eta^{(a)} \nonumber \\
&&\Delta_{ab} \rightarrow \Delta_{ab}+\eta^{(a)} -\eta^{(b)} \\
&&\Delta_m^{(a)} \rightarrow \Delta_m^{(a)} + k_a\eta^{(a)} \nonumber 
\end{eqnarray}

\subsection{Relation with the geometry}

We want to put forward the immediate coincidence of the mesonic expression for the volume
of the Sasaki-Einstein space $Y$ as discussed in  section \ref{sec:FeqHil},
\begin{equation}\label{HILtor}
  \text{Hil}(t_i; \mathcal{M}) \sim \frac{\mathrm{Vol}(a_i; Y)}{\epsilon^4} + \dots \; ,
\end{equation}
with the free energy of the field theory evaluated at large $N$.\footnote{
Related discussions appeared in \cite{Gulotta:2011si,Berenstein:2011dr}.
}
The free energy is a function of the 
conformal dimensions $\Delta_a$ of the fields $X_a$ and we can identify
\begin{equation} 
F\left( \Delta_a \right) = N^{3/2} \sqrt{\frac{2 \pi^6}{27 \, {\rm Vol}\left(a_i; Y \right)}} \,,
\label{eq:FVol}
\end{equation}
where $ \Delta_a = \sum_i P_{a i} \, a_{i} $, 
with $i$ running over the external perfect matchings and $P_{a i}$ being the matrix introduced in section
\ref{sec:TheToricDescription}.
According to the discussion in section \ref{sec:N2FieldTheory},
due to the gauge symmetries \eqref{flat} of $F$ 
we can identify $G-2$ baryonic symmetries which do not contribute to the free energy functional.
This is reflected by the invariance of the Hilbert series under the same symmetries, 
as we projected on the mesonic singlets.
A similar decoupling is well-known from the four-dimensional case \cite{Butti:2005vn}.

We identify several advantages when inferring the volumes from the Hilbert series.
First, (\ref{HILtor}) provides a fast and more direct way than \eqref{MSvol2} to obtain the geometrical 
informations of the volumes, without the need of mapping the $R$-charges of the PM 
with the volumes of the $5$-cycles as in \eqref{eq:DeltaofR}.
Moreover  we can compute the Hilbert series even in non-toric models,
where we cannot use the simple formulas \eqref{MSvol2} and \eqref{eq:Hilbert} anymore,
opening the way for a more general analysis as in \cite{Eager:2010yu}.

In the appendix \ref{sec:Zfunction} we also discuss the matching of the field theory free energy with
the geometrical $Z$-function at arbitrary Reeb vector.

\subsection{Examples}
\label{sec:FvectorExamples}

We now apply the discussion above to compute the free energy of some vector-like models. 
Our aim is to compare the localized quantity with the pole of the meson counting function introduced in section \ref{sec:FeqHil}.
In all our examples we find that
the result from the Hilbert Series and the large $N$ free energy coincide even before extremization.
Our results do not rely on the underlying symmetries enjoyed by the quiver gauge theories at the infrared fixed point
and generalize some of the results in \cite{Martelli:2011qj}.

We study the vector-like theories $\mathbb{C}\times \mathcal{C}$, 
$\widetilde{\text{SPP}}$ and  $\widetilde{{\cal C}/\mathbb{Z}_2}$.
The mesonic moduli space and the Hilbert Series of the first two models have already been
studied in \cite{Hanany:2008fj}, there
the tilded names are inherited from the four-dimensional theories which have the same quiver but
YM instead of CS interactions. The three theories are connected by an RG flow, 
which on the field theory side corresponds to giving a VEV to one of the scalar fields and then integrating it out.
This is reproduced on the gravity side by a partial resolution of the singularity, which can be
conveniently represented as removing an external point in the toric diagram. This is equivalent in
the geometric RG flow to blowing up a singularity, which in turn implies that the volume of
the manifold increases.
Hence, once established the relation $F^2\sim 1/\text{Vol}$, 
the decreasing of $F$ follows immediately in these cases, in agreement with the conjectured
$F$-theorem.

\begin{figure}
\centering
\subfigure{
\includegraphics[width=.4\textwidth]{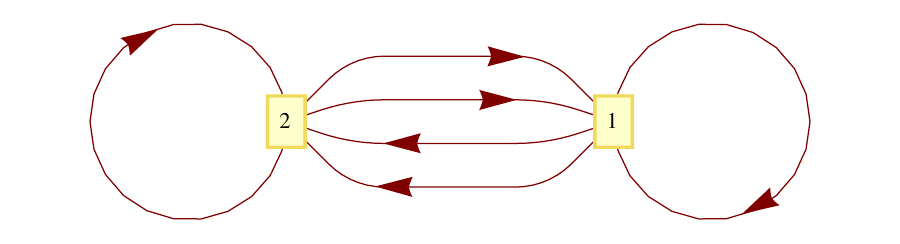}}
\hspace{0cm}
\subfigure{
\includegraphics[width=.4\textwidth]{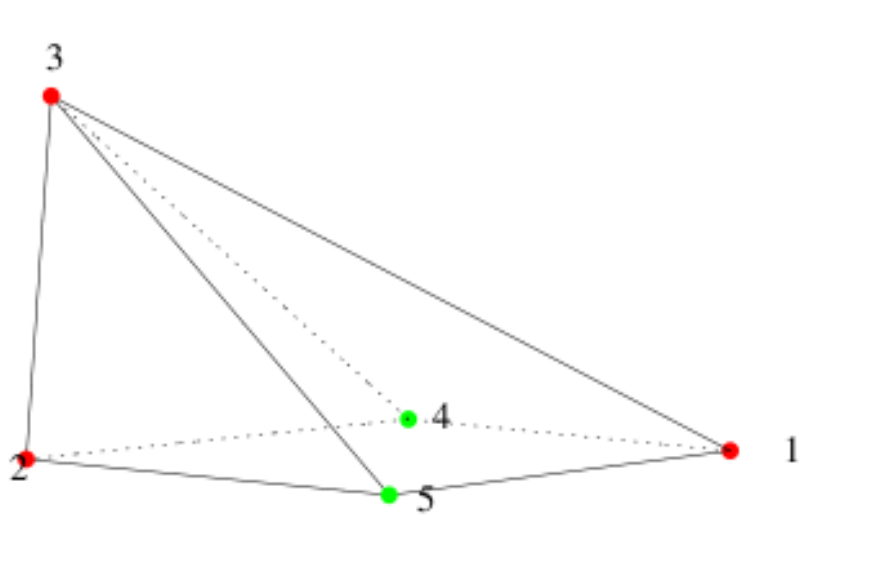}}
\caption{Quiver and toric diagram for $\mathbb{C} \times \cal C$.}
\label{fig:CxConif}
\end{figure}

\paragraph{$\mathbb{C} \times \mathcal{C}$.}
Consider a theory with gauge group $U(N)_k\times U(N)_{-k}$, 
two adjoints $\phi_i$ and two pairs $A_i, B_i$, $i=1,2$ of bifundamental fields in the $(\mathbf{N},\bar{\mathbf{N}})$
and $(\bar{\mathbf{N}},\mathbf{N})$, respectively,
as depicted in the quiver of figure \ref{fig:CxConif}.
The superpotential is
\begin{equation}
  W = \Tr \left(\phi_1 (A_1 B_2 - A_2 B_1) + \phi_2 (B_1 A_2 - B_2 A_1)  \right)
  \label{eq:WAB}
\end{equation}
and the moduli space is $\mathbb{C} \times \mathcal{C}$, where $\cal C$ is the
conifold.
Finding the exact superconformal $R$ symmetry requires, a priori,
an arbitrary choice of combining the $G+2=4$ abelian symmetries
(subjected to $R_{\text{trial}}[W]=2$) to parametrize the dimensions $\Delta$ 
and eventually finding the exact choice of $R$ by extremizing $F$.
We want to keep an eye on the correspondence and parametrize the dimensions 
in a way that allows for natural comparison with the geometry even before extremization.
To this end, we assign to each external perfect matching $p_i$ the charge $a_i$,
where we deliberately over-count global symmetries by the number of relations between the external $p_i$'s.
Since perfect matchings correspond to points in the toric diagram, given in figure \ref{fig:CxConif},
we can then directly incorporate the toric data of the moduli space.

The perfect matching matrix suggests the charge assignment 
\begin{equation}
\begin{split}
  &\Delta_{A_1} = a_1 + a_4 \qquad \Delta_{A_2} = a_2 + a_4 \qquad \Delta_{\phi_1} = a_3 \\
  &\Delta_{B_1} = a_1 + a_5 \qquad
\Delta_{B_2} = a_2 + a_5 \qquad \Delta_{\phi_2} = a_3
  \label{eq:ABa}
\end{split}
\end{equation}
where the marginality condition on the superpotential $\Delta(W)=2$ is reflected by $\sum_i a_i =2$.
Following the rules outlined in the previous section, we get the free energy functional
\begin{align}
  \frac{F[\rho,u,\mu]}{N^{3/2}} &= 
  \int \dd x \left( 
  \frac{k}{2\pi}\, \rho\, x\, u + 
  \rho^2 \left[ 
  P_{a_i} - a_3 (\pi(a_4-a_5) + u)^2 
   \right]  \right)
   -\frac{\mu}{2\pi} \left(\int \rho-1
   \right) \nonumber \\
  \shortintertext{where} 
   P_{a_i} &= - a_3 \pi^2 \left( (a_1 - a_2)^2 - (a_3-2)^2\right)
  \label{FABx}
\end{align}
and we defined $u=u(x)\equiv y_1(x) -y_2(x)$.
Note that we have included the monopole charge $\Delta_m$ not via a topological term in the free energy functional,
but via $a_{4/5} \rightarrow a_{4/5} \pm 2\Delta_m $,
corresponding to the direction in the abelian gauge space which is broken to $\mathbb{Z}_k$.
This can be done by shifting $y^a \rightarrow y^a - \Delta_m/k^a$.
The free energy functional is extremized for
\begin{equation}
 \rho = 
    \begin{cases}
        \frac{\mu +2\, k\, \pi \, x\, A_2}{8 \pi ^3 \left(A_1-A_2\right) \phi \left(A_2+B_2\right)}
	& , \; \; -\frac{\mu }{2 k \pi  A_2} < x < -\frac{\mu }{2 k \pi  A_1} \\
       \frac{\mu /\pi +k\,x \left(B_1-B_2\right)}{4 P_{a_i}} 
	& , \; \; -\frac{\mu }{2 k \pi  A_1} <x < \frac{\mu }{2 k \pi  B_1} \\
        \frac{\mu -2 k\, \pi \, x B_2}{8 \pi ^3 \left(A_1-A_2\right) \phi \left(A_2+B_2\right)}
	& ,  \: \; \; \; \; \frac{\mu }{2 k \pi  B_1} <x < \frac{\mu }{2 k \pi  B_2}
    \end{cases} \; ,
   \label{eq:ABrho}
\end{equation}
where, without loss of generality, we assumed that $a_1>a_2$ and $a_4>a_5$.
Furthermore, we used (\ref{eq:ABa}) and for the ease of notation we denoted $\psi \equiv \Delta_{\psi}$ for a field $\psi$.
In the outer regions of (\ref{eq:ABrho}), $u$ is frozen to $u_{min} = -2 \pi (a_2+a_4)$
and $u_{\text{max}} = 2\pi (a_2+a_5)$, respectively.
In the middle region we find
\begin{equation}
 u(x) = \frac{k P_{a_i}x}{a_3 \left(\mu +k \pi  x \left(a_4-a_5\right)\right)}-\pi \left(a_4-a_5\right) \,.
\end{equation}
The Lagrange multiplier $\mu$ is fixed by $\int \rho = 1$ and the free energy finally reads
\begin{align}
  \left( \frac{F}{N^{3/2}} \right)^2 
  = \frac{32 \pi ^2 k \, a_3 \left(a_1+a_4\right) \left(a_2+a_4\right) \left(a_1+a_5\right) \left(a_2+a_5\right)}{9 \left(2-a_3\right)}
  \label{eq:ABFfinal}
\end{align}
We want to compare this to the Hilbert series. 
From the toric data in figure \ref{fig:CxConif} and $\eqref{eq:ABtoric}$, we read off the monomial weights
\begin{equation}
  (Z_i)=(z,z,1,z^{-1},z^{-1}) \,.
\end{equation}
We can then compute the Hilbert series $\eqref{eq:Hilbert}$, 
\begin{equation}
  \mathrm{Hil}(Y_{\mathbb{C}\times \mathcal{C}}; \,t_i) 
  = \oint \frac{\dd z}{2\pi i z} \frac{1}{(1-t_1 z)(1-t_2 z)(1-t_3 )(1-t_4/ z)(1-t_5/ z)} \,,
  \label{eq:ABHil}
\end{equation}
whose pole for $t_i = e^{-2 \varepsilon a_i} \rightarrow 1$ indeed reveals $1/F^2$ from $\eqref{eq:ABFfinal}$.

That this is a good description of the mesonic moduli space might look puzzling,
when only counting parameters.
As mentioned above, there are generically more $a_i$'s, 
namely $(G+2+\text{(\# of relations on the external pm's)})$
\footnote{
This redundancy amongst the external perfect matchings   
originates  from the splitting of 
points in the parent 2d diagram, which have a 
multiplicity \cite{Hanany:2008fj}.
These points may sit on the perimeter or in the internal of the diagram.
Depending on this, the new external points of the 3d diagram may not be in $1:1$ correspondence 
with the ($G+2$ many) global symmetries of the CFT$_3$.
This is opposed to four dimensional theories, 
where the number of the external points is always identical with the number of non-anomalous global symmetries.
When going to 3d, the anomalies disappear and all $G+2$ global symmetries are physical.
Pick's theorem relates the number of these symmetries to the properties of the 2d toric diagram,
$
G+2 = \text{Perimeter} + 2 (\text{Internal Points}) \, .
$
We see that precisely in the cases in which all internal points split, 
the number of external points of the split$3$d diagram is $G+2$.
Else, there are extra points coming from split points on the perimeter.
This is the case for the vector-like theories discussed in this section.
We marked  the split points by green dots. 
}
then there are mesonic symmetries, namely $4$.
The key observation is though, that the $a_i$'s appear only in combinations
of meson charges, hence modulo baryonic and spurious symmetries.
In our example this is particularly easy since, given $G=2$, 
there are no baryonic symmetries in the game.
We do, nevertheless, identify the non-physical, spurious symmetry
\[
a_{1/2} \rightarrow a_{1/2} + s \, , \qquad a_{4/5} \rightarrow a_{4/5} - s \, ,
\]
which reflects the relation $p_1+p_2=p_4+p_5$ 
of the perfect matchings and reduces the number of independent $a_i$'s to 4.
These can in principle be mapped to the charges under the Cartan-part of the global symmetry
\[
SU(2)_1\times SU(2)_2\times U(1)_1 \times U(1)_2 \, .
\]
In absence of baryonic symmetries
the bifundamental fields by themselves are mesonic operators, 
and their dimensions appear in the final result for $F$.

We conclude this example by observing that
in \cite{Jafferis:2011zi} the authors discussed a dual phase of this theory,
which involves fundamental flavor fields.
Upon the identifications of the PM
the two expressions for the free energy coincide.

\paragraph{$\widetilde{SPP}$.}
\begin{figure}
\centering
\subfigure{
\includegraphics[width=.3\textwidth]{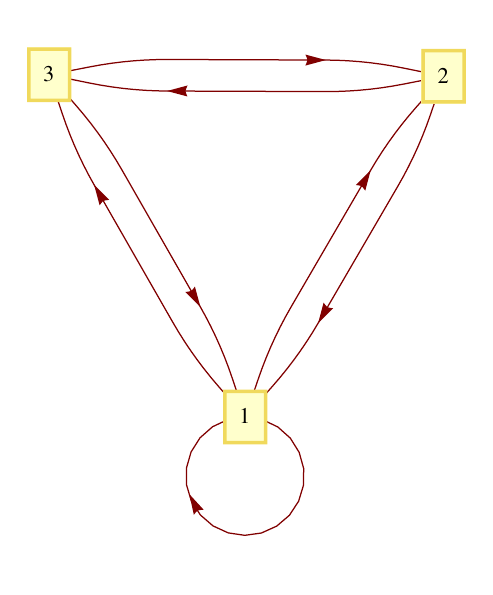}}
\hspace{1.5cm}
\subfigure{
\includegraphics[width=.5\textwidth]{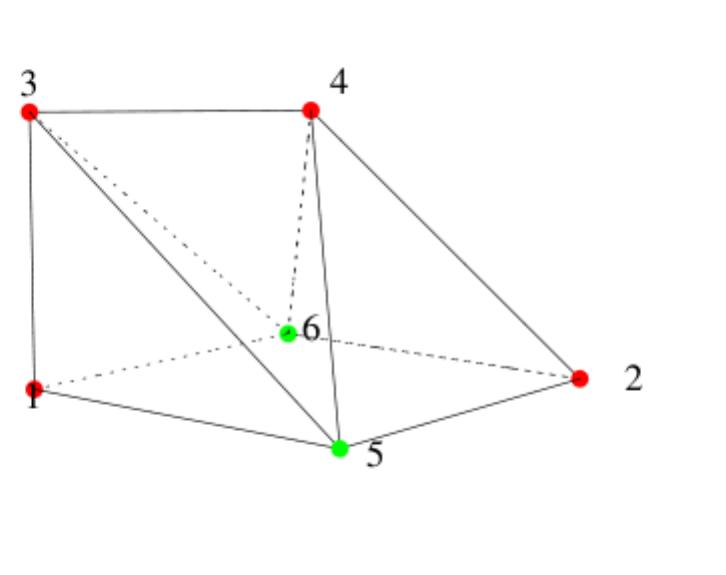}}
\caption{Quiver and toric diagram for $\widetilde{SPP}$. The diagram is plotted for CS levels $(2,-1,-1)$.}
\label{fig:SPP}
\end{figure}
Next, we want to study the quiver in figure \ref{fig:SPP} with gauge group
$U(N)_1\times U(N)_2 \times U(N)_3$, one adjoint $\phi$ of the $U(N)_1$, and three pairs $A_i,B_i,C_i$ of (anti) bifundamentals in the
$(\mathbf{N},\mathbf{\bar{N}},1),(1,\mathbf{N},\mathbf{\bar{N}}),(\mathbf{\bar{N}},1,\mathbf{N})$ representation of the gauge group, respectively,
and Chern-Simons couplings $(-k_2-k_3,k_2,k_3)$. The superpotential reads
\begin{equation}
  W = \phi (A_1 A_2 - C_2 C_1) - A_2 A_1 B_1 B_2 + C_1 C_2 B_2 B_1 \, .
  \label{eq:ABCW}
\end{equation}
As special cases, the family includes $\widetilde{SPP_{-211}}$ and $D_3 = \widetilde{SPP_{1-10}}$.
From the perfect matching matrix, we again infer the charge assignment \cite{Hanany:2008fj}
\begin{equation}
  \Delta_{A_i} = a_i + a_{i+4}, \; \, \Delta_{B_i} = a_{i+2}, \;\, \Delta_{C_i} = a_i + a_{7-i} , \; \, \Delta_{\phi} = a_3+a_4\,,
  \label{eq:ABCa}
\end{equation}
where the six $a_i$'s include one redundancy and one baryonic direction.
For the ease of notation, let us introduce the combinations
\begin{equation}
  A_- = \Delta_{A_1} - \Delta_{A_2}, \;\;\; B_- = \Delta_{B_1} - \Delta_{B_2}, \;\; \;C_- = \Delta_{C_1} - \Delta_{C_2} , \;\;\; B_+ = \Delta_{B_1} + \Delta_{B_2} \,,
  \label{eq:ABCDeltamp}
\end{equation}
a convenient parametrization for solving the saddle point equations.
The free energy functional is given by
\begin{align}
  \frac{F[\rho,u,v,\mu]}{N^{3/2}} &=
-\frac{\mu  (\int \rho -1)}{2 \pi } + \int \Bigg[
-\frac{x \rho  \left(k_2 u_1-k_3 u_3\right)}{2 \pi } \nonumber \\
+&\frac{\rho^2}{2} \left(P-B_+ (\pi A_-+u_1)^2- (2-B_+) (\pi  B_--u_1-u_3)^2-B_+ (\pi  C_-+u_3)^2\right) \Bigg] \, \nonumber,\\
\shortintertext{with}
P&= \pi ^2 \left(-4+B_+\right) \left(-2+B_+\right) B_+ \,.
\label{eq:ABCF}
\end{align}
Here $u=y^1-y^2$ and $v=y^3-y^1$.
For arbitrary CS levels and R-charges,
the eigenvalue distribution is generically divided in five regions.
We refrain from writing down the explicit functions $u(x), v(x)$ and $\rho(x)$,\footnote{
Beyond the central region where (\ref{eq:ABCF}) is extremized, 
there's a middle region with constant
$u$ or $v$ (depending on relations amongst the $k_a$'s and the $a_i$'s).
Finally, in the outer regions, both $u$ and $v$ are constant and $\rho$ is eventually becoming zero.
}
since their expressions are cumbersome and not illuminating.
We have computed $F$ with arbitrary levels $k_2$ and $k_3$,
where we had to made a choice on the relative sign.
We skip the general expression because it is too cumbersome, and we
focus on two specific examples.
Nevertheless, we checked the agreement with the geometry at arbitrary levels $k_i$.
Consider $\widetilde{SPP_{-211}}$
\begin{align}
  \scriptstyle \left(\frac{F}{N^{3/2}}\right)^2=\frac{\pi ^2 \left(4+A_--C_--2 B_+\right) \left(4+2 A_-+B_--B_+\right) \left(4+B_-+2 C_--B_+\right) B_+ \left(4-2 A_--B_--B_+\right) \left(4-B_--2 C_--B_+\right) \left(4-A_-+C_--2 B_+\right)}{9 \left(128+2 A_-^2 \left(-4+B_+\right)+2 C_-^2 \left(-4+B_+\right)-112 B_+-B_-^2 B_+-2 B_- C_- B_++32 B_+^2-3 B_+^3-2 A_- \left(4 C_- \left(-2+B_+\right)+B_- B_+\right)\right)} \nonumber
\end{align}
and $D_3 = \widetilde{SPP_{1-10}}$
\begin{align} 
  \left( \frac{F}{N^{3/2}} \right)^2= \frac{1}{9} \pi ^2 \left(2-B_--C_-\right) \left(2+B_-+C_-\right) \left(2-A_--B_+\right) \left(2+A_--B_+\right) B_+ \nonumber\\
   \label{FD3final}
\end{align}
Note that upon $\eqref{eq:ABCa}$, these are expressions in terms of the $a_i$'s. 
The monopole charge is included along
\begin{align}
  &\delta a_1 \sim k_3 \Delta _m,& &\delta a_3 \sim -2 \left(k_2+k_3\right) \Delta _m, & & \delta a_5\sim\left(k_2-k_3\right) \Delta _m,\nonumber\\
  &\delta a_2\sim k_2 \Delta _m,& &\delta  a_4\sim\left(k_2+k_3\right) \Delta _m, & &\delta a_6\sim\left(k_3-k_2\right) \Delta _m \nonumber ,
\end{align}
which corresponds to the direction in gauge space parallel to the $k$'s.
The overcounting is reflected by the spurious symmetry
\[
\delta a_{1/2} \sim s \, , \qquad \delta a_{3/4} \sim - s \, 
\]
and also the contribution of the baryonic symmetry $k_3 U(1)_2 - k_2 U(1)_3$, 
\begin{align*}
   &\delta a_1 \sim -b k_2,&  &\delta a_3\sim b \left(k_2-k_3\right),&   &\delta a_5\sim b \left(k_2+k_3\right), \\
   &\delta a_2 \sim b k_3,&  & \delta a_4 \sim 0,& & \delta a_6\sim -b \left(k_2+k_3\right)\,,
\end{align*}
is indeed a symmetry of $F$.

For the Hilbert series, we extract the weights of the quotient from the toric data in figure \ref{fig:SPP} and $\eqref{eq:ABCtoric}$,
\begin{equation}
  (Z_i)=(w,w z^{-k_2-k_3},z^{-k_2-k_3},z^{k_2+k_3},w^{-1} z^{k_3},w^{-1} z^{k_2}) \,, 
\end{equation}
from which we compute 
\begin{multline} \label{eq:HilSPP}
  \mathrm{Hil}(Y_{\rm SPP}; \,t_i) = \\ \oint \frac{\dd z \, \dd w}{(2\pi i)^2 z w} 
  \frac{1}{(1-t_1 w)(1-t_2 w / z^{k_2+k_3})(1-t_3/ z^{k_2+k_3})(1-t_4 z^{k_2+k_3})(1-t_5 z^{k_3}/w)(1-t_6 z^{k_2}/w)} \,.
\end{multline}
The $t_i = e^{-2 \varepsilon a_i} \rightarrow 1$ pole of $\eqref{eq:HilSPP}$ reproduces the free energy,
where one again has to make choices on the signs of the $k_i$'s.

\paragraph{$\widetilde{\mathcal{C}/\mathbb{Z}_2}$.}
\begin{figure}
\centering
\subfigure{
\includegraphics[width=.3\textwidth]{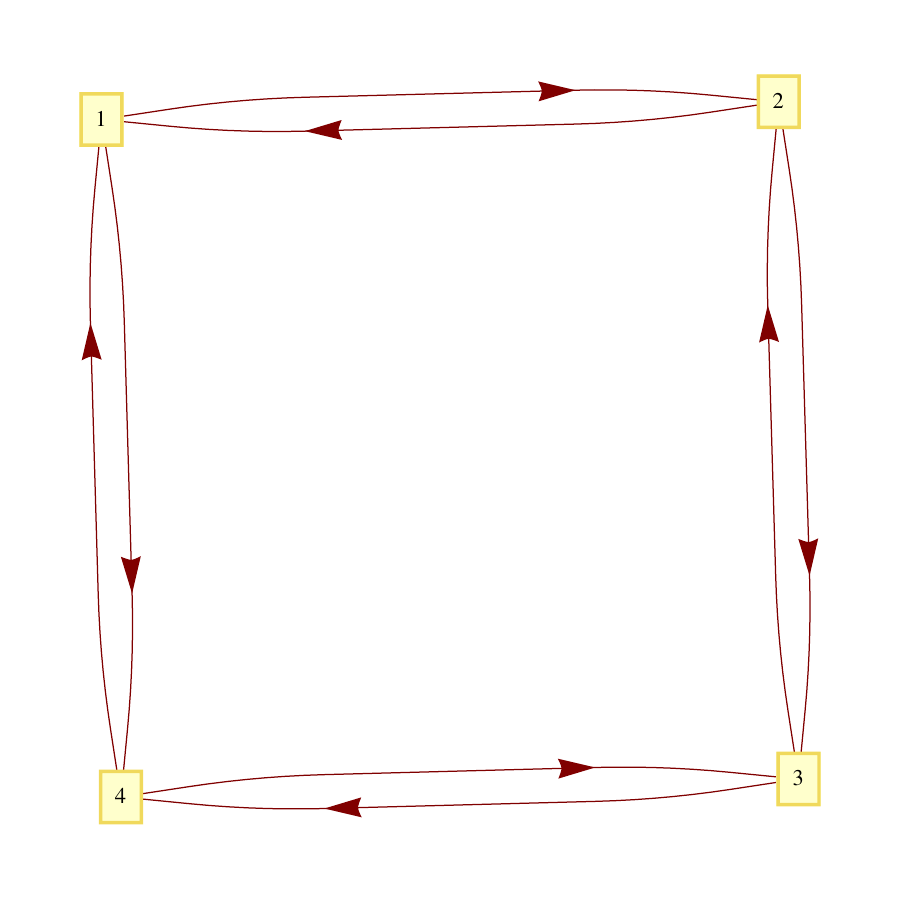}}
\hspace{1.5cm}
\subfigure{
\includegraphics[width=.25\textwidth]{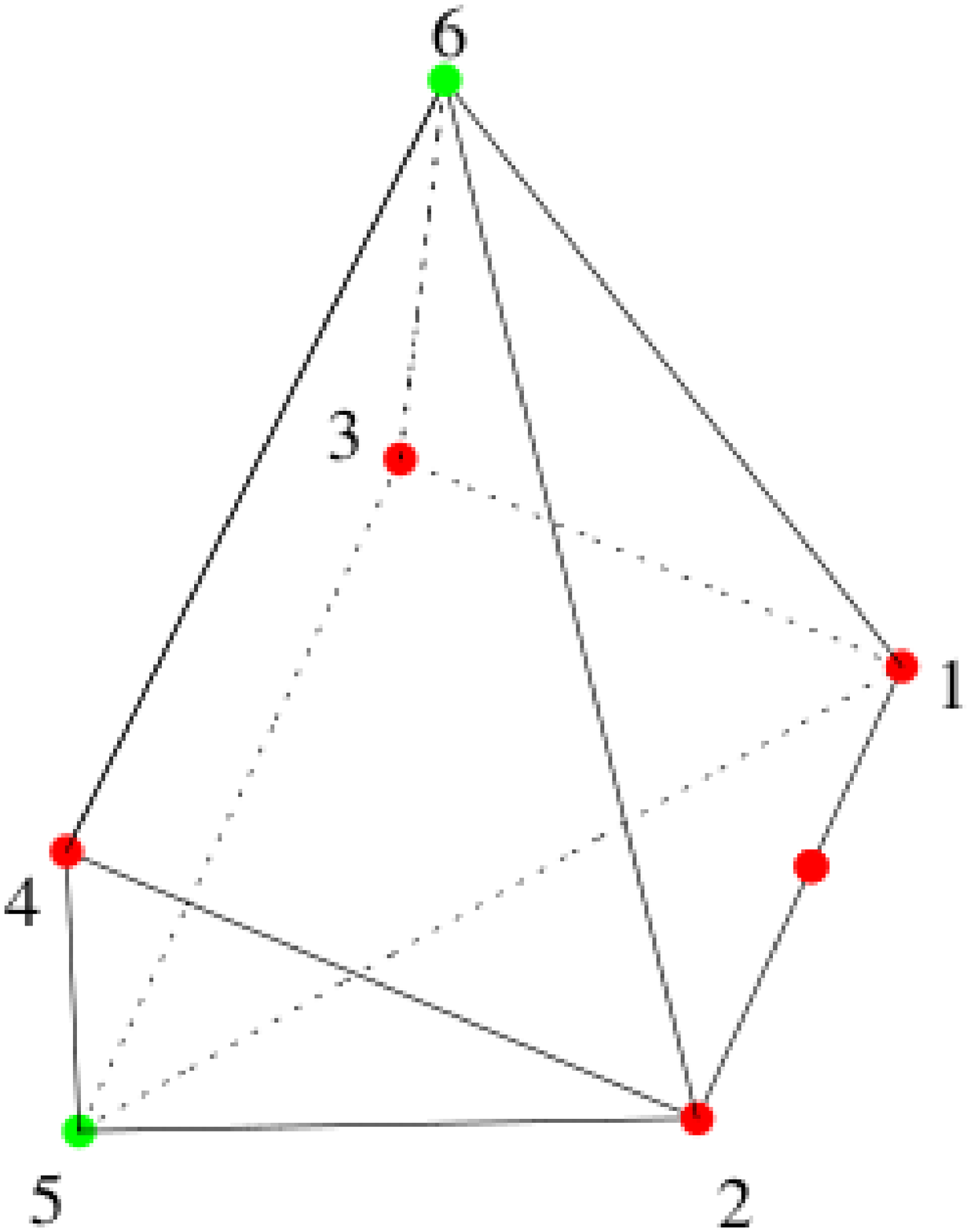}}
\caption{Quiver and toric diagram for $\widetilde{\mathcal{C}/\mathbb{Z}_2}$.}
\label{fig:GenConif}
\end{figure}
The generalized conifold with $\mathcal{N}=3$ supersymmetry has been studied in \cite{Herzog:2010hf},
here we do not want to assume $\mathcal{N}=3$ supersymmetry and consider the quiver as a $\mathcal{N}=2$ model, i.e. we assign arbitrary $R$ charges
to the fields.
The field content is shown in figure \ref{fig:GenConif}, the CS couplings are $(k,k,-k,-k)$ and
we parametrize the R-charges of the fields corresponding to the perfect matchings
\begin{align}
  \Delta_{A_i/C_i} = a_i \, , \; \; 
  \Delta_{B_i} = a_{i+2} + a_{i+4}  \, ,\; \;
  \Delta_{D_i} =  a_{i+2} + a_{7-i} \,.
  \label{eq:ABCDa}
\end{align}
There is no redundancy but two baryonic symmetries,
which are no actual degrees of freedom in the free energy.
The eigenvalue distribution is divided in five parts, again we refrain from giving all the formulae and just present the result
\begin{equation}
  \left( \frac{F}{N^{3/2}} \right)^2 =
\frac{2 k \pi ^2  A_+ \left(4-\left(A_-+B_-\right)^2\right) \left(4-\left(A_-+D_-\right)^2\right) 
\left(\left(B_--D_-\right)^2-4 \left(2-A_+\right)^2\right)}
{9 \left(A_+\left(4+A_-^2+B_- D_-+A_- \left(B_-+D_-\right)\right) +\left(B_--D_-\right)^2-16\right)} \, ,
\label{eq:genConF}
\end{equation}
where we used a similar rewriting as in $\eqref{eq:ABCDeltamp}$.

From the toric diagram in figure \ref{fig:GenConif} and $\eqref{eq:ABCDtoric}$,
we read off the charge matrix for the Hilbert series
\begin{multline}
  \mathrm{Hil}(Y_{\rm gen. Con.}; \,t_i) = \\ \oint \frac{\dd z \, \dd w}{(2\pi i)^2 z w} 
  \frac{1}{(1-t_1 w)(1-t_2 /w )(1-t_3/ (w z))(1-t_4 w / z)(1-t_5 z)(1-t_6 z)} \,.
\end{multline}
Its pole for $t_i = e^{-2 \varepsilon a_i} \rightarrow 1$ immediately reproduces $\eqref{eq:genConF}$.\\

Let us comment on the RG flow between the three theories discussed so far.
We can follow the flow between the fixed points
$\widetilde{\mathcal{C}/\mathbb{Z}_2} \rightarrow \widetilde{SPP} \rightarrow \mathbb{C}\times \mathcal{C}$
by partially resolving the singular spaces.
This corresponds to removing points in the toric diagram \cite{Franco:2008um}.
More explicitly, upon removing point $1$ and one of the internal points in figure \ref{fig:GenConif},
we obtain the diagram of figure \ref{fig:SPP}, up to renaming.
In the field theory, this corresponds to giving a VEV and integrating out  $A_1=p_1 p_7$.
Note that $p_7$ is an internal perfect matching, which is the reason we have omitted it in the discussion so far.
The groups $U(N)_{k_1}$ and $U(N)_{k_2}$ are identified to $U(N)_{k_2+k_3}$ and $A_2$ becomes the adjoint field in $\widetilde{SPP}$. 
If we now in figure \ref{fig:SPP} remove also point $4$, 
we end up with the diagram of $\mathbb{C}\times \mathcal{C}$ in figure \ref{fig:CxConif}, modulo relabeling the points.
In the field theory this is achieved by higgsing $B_2=p_4$.

\paragraph{ABJM$/\mathbb{Z}_2$.}
\begin{figure}
\centering
\subfigure{
\includegraphics[width=.3\textwidth]{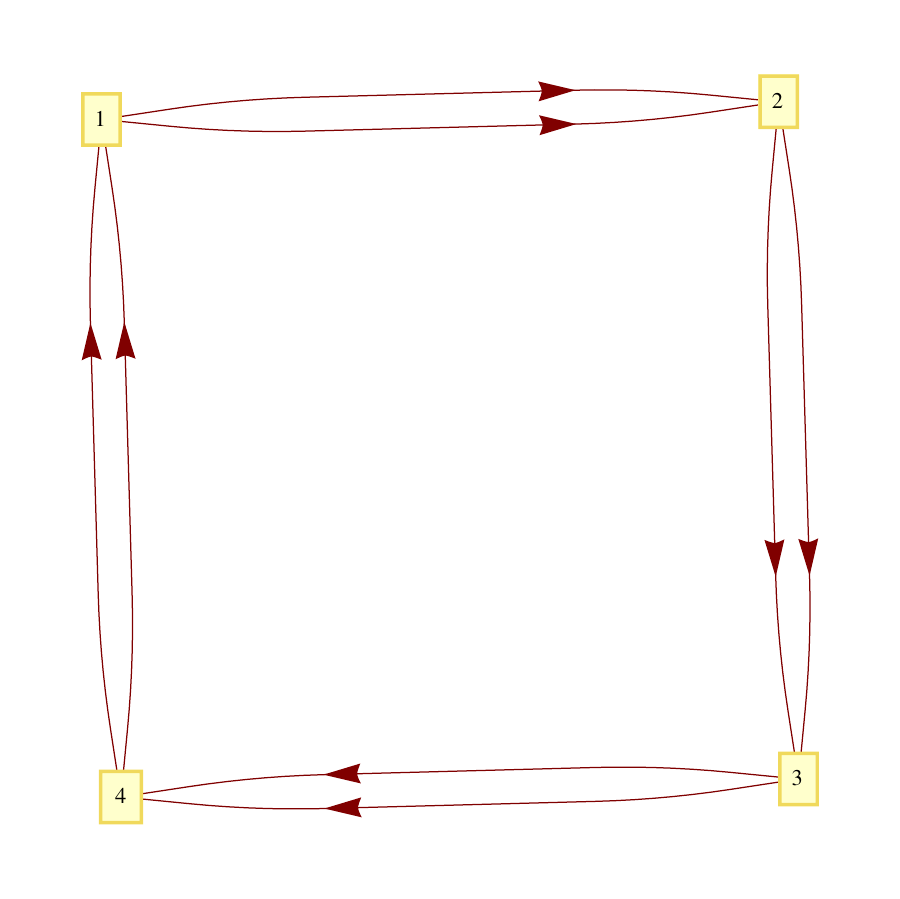}}
\subfigure{
\includegraphics[width=.33\textwidth]{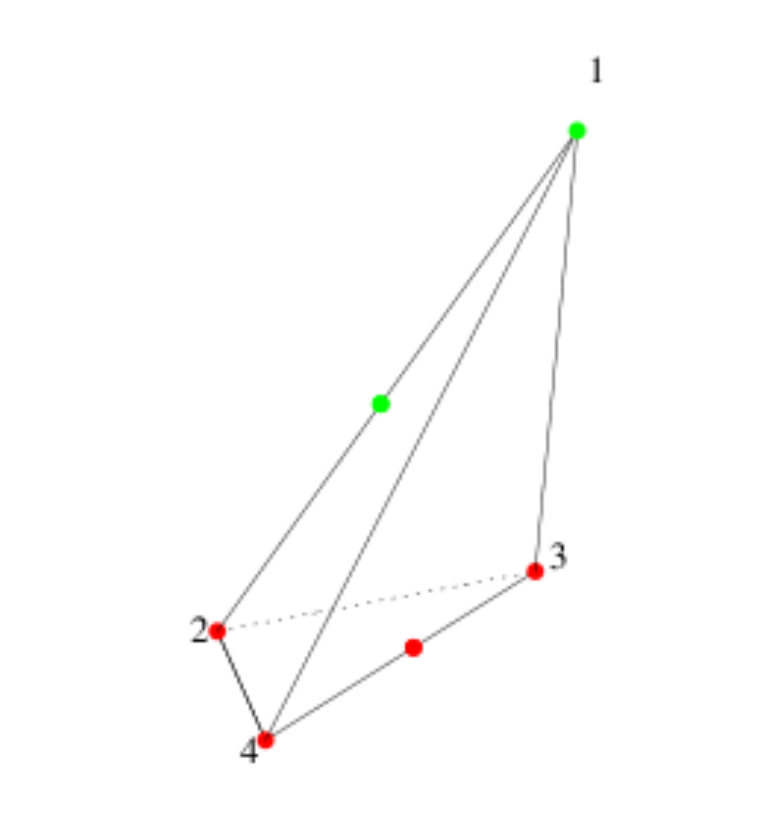}}
\caption{(Left) Quiver for $\widetilde{\mathbb{F}_0}$. According to the choice of the CS levels, it
gives several theories studied in the paper. (Right) Toric diagram for ABJM$/\mathbb{Z}_2$, which
corresponds to CS levels $(k,-k,k,-k)$.}
\label{fig:quivABJM}
\end{figure}
We consider the theory with product gauge group $\prod_{i=a}^4 U(N)_a$, Chern-Simons levels $(k,-k,k,-k)$
and four pairs of bifundamental fields $A_i,B_i,C_i,D_i$ as shown in figure \ref{fig:quivABJM}.
At a first look, the theory seems chiral and it is not clear how the long-range forces vanish without modifications.
Taking into account the symmetry of the quiver, though, 
we find that the contribution to the long range forces coming from $A/B$ cancels with that of $C/D$, respectively.
In fact, the theory can be seen as a $\mathbb{Z}_2$ quotient of ABJM,
effectively being vector-like and having a saddle point solution following the ansatz used so far.
We assign to the fields charges under the perfect matchings,
\begin{align}
  \Delta_{A_i/C_i} = a_i \, , \; \; 
  \Delta_{B_i/D_i} = a_{i+2} \, ,\; \;
  \label{eq:ABJMa}
\end{align}
where the affiliation to ABJM is manifest: Both baryonic directions are killed by the $\mathbb{Z}_2$ flip symmetry
of the quiver and we are left with the $4$ mesonic charges only.
Imposing the symmetry $y^1=y^3$ and $y^2=y^4$, 
makes the orbifold of ABJM obvious even at the level of the free energy functional.
As a solution we find consequently 
\begin{equation}
  \left( \frac{F}{N^{3/2}} \right)^2 =
\frac{128}{9} k \pi ^2 a_1 a_2 a_3 a_4 \,.
\label{eq:ABJMF}
\end{equation}
The toric diagram is given in figure \ref{fig:quivABJM} and $\eqref{eq:ABJMtoric}$.
Modulo a discrete $\mathbb{Z}_2$, the Hilbert series is trivial
\begin{equation}
  \mathrm{Hil}(S^7; \,t_i) 
  = \frac{1}{(1-t_1)(1-t_2)(1-t_3)(1-t_4)} \,,
\end{equation}
matching with $\eqref{eq:ABJMF}$ as $t_i = e^{-2 \varepsilon a_i} \rightarrow 1$. 

\section{Seiberg duality in vector-like theories}
\label{sec:Seiberg}

In this section we show that the large $N$ free energy preserves the 
rules of Seiberg duality for vector like gauge theories
worked out in \cite{Giveon:2008zn}.
First of all, we review the rules of Seiberg duality in three dimensional 
vector like CS matter theories with product groups, and their relation with toric duality.
In three dimensions a vector multiplet can have either a YM or a CS  term in the action. In the first case the 
theory is similar to the four dimensional parent  but the rules of duality cannot be extended straightforwardly. Indeed, the vector 
multiplet has an additional scalar coming from the dimensional reduction which modifies the moduli space. 
As a consequence it was observed in \cite{Aharony:1997bx} that the rules of Seiberg duality 
are modified by adding new gauge invariant degrees of freedom in the dual magnetic theory,
which take into account the extra constraints on the moduli space.
On the other hand, YM-CS (or even CS) theories do have a dual description with the same field content as their four dimensional parents.
The only difference is on the gauge group. 
Indeed for CS SQCD with $U(N)_k$ gauge group and $N_f$ pairs of $Q$ and $\tilde Q$,
the dual field theory has $U(N_f+|k|-N_c)_{k}$  gauge group, as shown in \cite{Giveon:2008zn}.
The partition function has already been used to 
check this extension of  CS $\mathcal{N}=2$ Seiberg duality in three dimensions
in \cite{Niarchos:2011sn,Willett:2011gp,Kapustin:2011gh,Morita:2011cs,Benini:2011mf,Kapustin:2011vz,Dolan:2011rp}.

One may then wonder if the same rules can be extended to more complicated gauge theories, like the ones related by AdS/CFT to the motion
of M$2$ branes on CY$_4$.  The first generalization of Seiberg like dualities on CS quiver gauge theories  appeared in \cite{Aharony:2008gk} 
for the ABJM model. It was observed that the field content transforms as in $4d$ while the gauge group transforms as
\begin{equation}
 U(N)_k\times U(N+M)_{-k} \rightarrow U(N)_{-k} \times U(N-M)_k
\end{equation}
Differently from the four dimensional case, also the gauge group spectator
feels the duality, since its CS level is modified.
The above rule can be derived by looking at the system of  branes
engineering the gauge theory.  This consists of a stack of $N$ D$3$ 
on a circle and two pairs of $(1,p)$ branes orthogonal to them.
Moreover $M$ fractional D$3$ branes on a semicircle connecting the fivebranes are added.
By moving the fivebranes on the circle and by applying the s-rule \cite{Hanany:1996ie}, when one stack of $(1,p)$ branes
crosses the other, the rule above is derived.

It is then natural to extend these ideas to theories with a higher number of gauge groups.
When these theories can be described as a set of $(1,p_i)$ and D$3$ branes on a circle,
they are the extension of the four dimensional $L^{aba}$ gauge theories 
\cite{Benvenuti:2005ja,Franco:2005sm,Butti:2005sw}. They consist of a product of
gauge groups $U(N_i)$ with bifundamentals and adjoints (the presence of the adjoint 
is related to the choice of the angles between the $(1,p_i)$ fivebranes).
In absence of an adjoint field on the node $N_i$ 
the interaction in the superpotential is  $W_i=X_{i-1,i}X_{i,i+1}X_{i+1,i}X_{i,i-1}$
while if there is an adjoint  on $N_i$ we have
$W_i=X_{i-1,i}X_{i,i}X_{i,i-1 } -X_{i+1}X_{i,i}X_{i,i+1}$. The signs as in four dimensions alternate
between $+$ and $-$.

Even in this case the duality rules are found by exchanging the $(1,p_i)$ and the $(1,p_{i+1})$ 
fivebranes. The final rule is
\begin{eqnarray} \label{SD}
U(N_{i-1})_{k_{i-1}}& \rightarrow & U(N_{i-1})_{k_{i-1}+k_i}\nonumber \\
U(N_{i})_{k_{i}} ~~~~&\rightarrow& U(N_{i-1}+N_{i+1}+|k_i|-N_i)_{-k_i}\\
U(N_{i+1})_{k_{i+1}} &\rightarrow &U(N_{i+1})_{k_{i+1}+k_i}\nonumber
\end{eqnarray}
while the matter field content and the interactions transform as in four dimensions.

\subsection{Matching the free energy}

In this section we provide the rules 
for the action of Seiberg duality (\ref{SD}) on the eigenvalues
of non chiral theories
and we show that the free energy matches even before the 
large $N$ integrals are performed.
Consider a duality on the $i$-th node.
The CS level of this group becomes $-k_i$ and
the imaginary part of this eigenvalue $y_i$
becomes $-y_i$.
Moreover the CS levels $k_{i\pm1}$ (here we just refer to
necklace quivers) become $k_{i\pm 1}+k_i$.
This rule and the constraint that the sum of the CS level is 
vanishing provide the duality action on the eigenvalues. 
We have
\begin{eqnarray} \label{rulesy}
y_{i+1} \rightarrow \tilde y_{i-1}= -y_{i-1} \quad &&\quad y_{i-1} \rightarrow  \tilde y_{j-1} =  -y_{i+1} \nonumber \\
y_{j} \rightarrow \tilde y_{j}=y_j - y_{i-1}-y_{i+1} &&\quad j \neq i,i\pm 1 
\end{eqnarray}

Note that the shift in the rank of the gauge group has a subleading effect at large $N$.
This apparently trivial statement is subtle,
since naively one finds new fundamental-like terms scaling like $N^{3/2}$, which descend from the $N^{5/2}$ contributions to $F$. 
To see this, let us consider a shift of the $i$-th rank by $\delta N$ and collect the additional contributions to the free energy 
following (2.9) of \cite{Jafferis:2011zi}.
At $N^{3/2}$, one has extra contributions $\rho \,x \,\delta N$ from the gauge sector,
$\rho \, x\, \delta N  /2(\Delta_{j,i} -1 + y^{j}/4\pi)$ from each incoming and
$\rho \, x\, \delta N  /2(\Delta_{i,j} -1 - y^{j}/4\pi)$ from each outgoing matter field, where $j=i\pm 1$.
We see that the net contribution cancels for the non-chiral theories at hand.
The dependency on the $y$'s drops out due to the vectorial nature of the quiver and 
the $y$-independent part is the anomaly cancellation of the $4d$ parent,
which has already been used in the treatment of the long-range forces.

In the case of a $\widetilde{L^{aba}}_{k_i}$ theory,
we  distinguish between the duality action on the classical
term of the free energy (\ref{integral}) and the one on the  loop contribution (\ref{oneloop}).
By supposing that the duality acts  on the $i$-th node,  the CS levels transform as in (\ref{SD}) while
the sum in the integral (\ref{integral}) becomes
\begin{equation}
\sum_{a=1}^{G} k_a y_a \rightarrow - k_i y_i +k_{i-1} y_{i-1}+ k_{i+1} y_{i+1} +\sum_{a\neq i,i\pm 1} k_a y_a
\end{equation}
By applying  (\ref{SD}) this last formula becomes
\begin{equation}
\sum_{a=1}^{G}  k_a y_a \rightarrow 
k_i \tilde y_i \sum_{a\neq i,i\pm 1} k_a \tilde y_a - \tilde y_{i-1} \sum_{b\neq i-1} k_b - \tilde y_{i+1} \sum_{c\neq i+1} k_c 
=
\sum_{a=1}^{G} k_a \tilde y_a
\end{equation}
The second term is the one loop contribution coming from the vector and the matter fields. 
In the non  chiral case of $\widetilde{L^{aba}}_{k_i}$ theories this contribution is 
\begin{equation}
F_{1L} =\sum_{a,b} F_{1L}^{a,b}
\end{equation}
where  $F_{1L}^{a,b}$ has been defined in (\ref{oneloop})
and 
the sum extends to the pairs of bifundamentals (a,b) and adjoint fields $(a,a)$ (counted twice).
We are going to show that the rules (\ref{SD}) leave $F_{1L}$ invariant.

\begin{figure}
\begin{center}
\includegraphics[width=12cm]{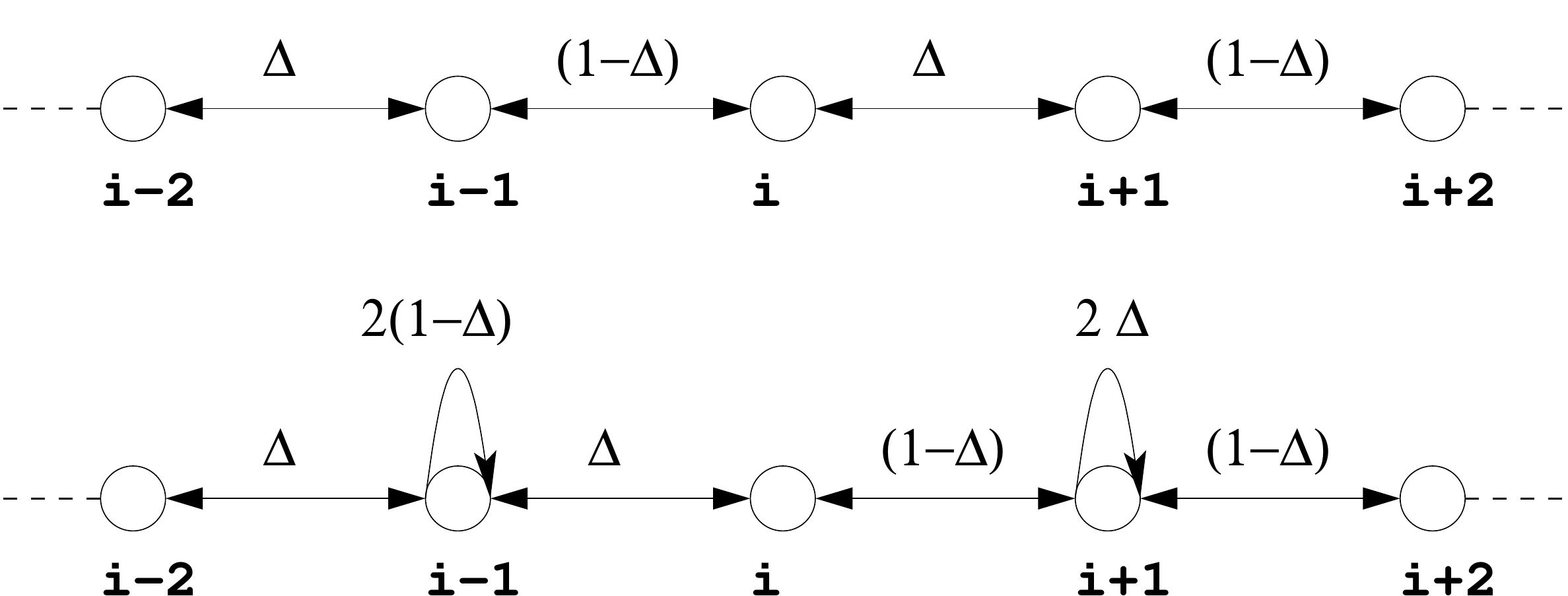}
\end{center}
\caption{Quiver diagram for the first type of the Seiberg duality}
\label{ciccio1}
\end{figure}

\begin{figure}
\begin{center}
\includegraphics[width=12cm]{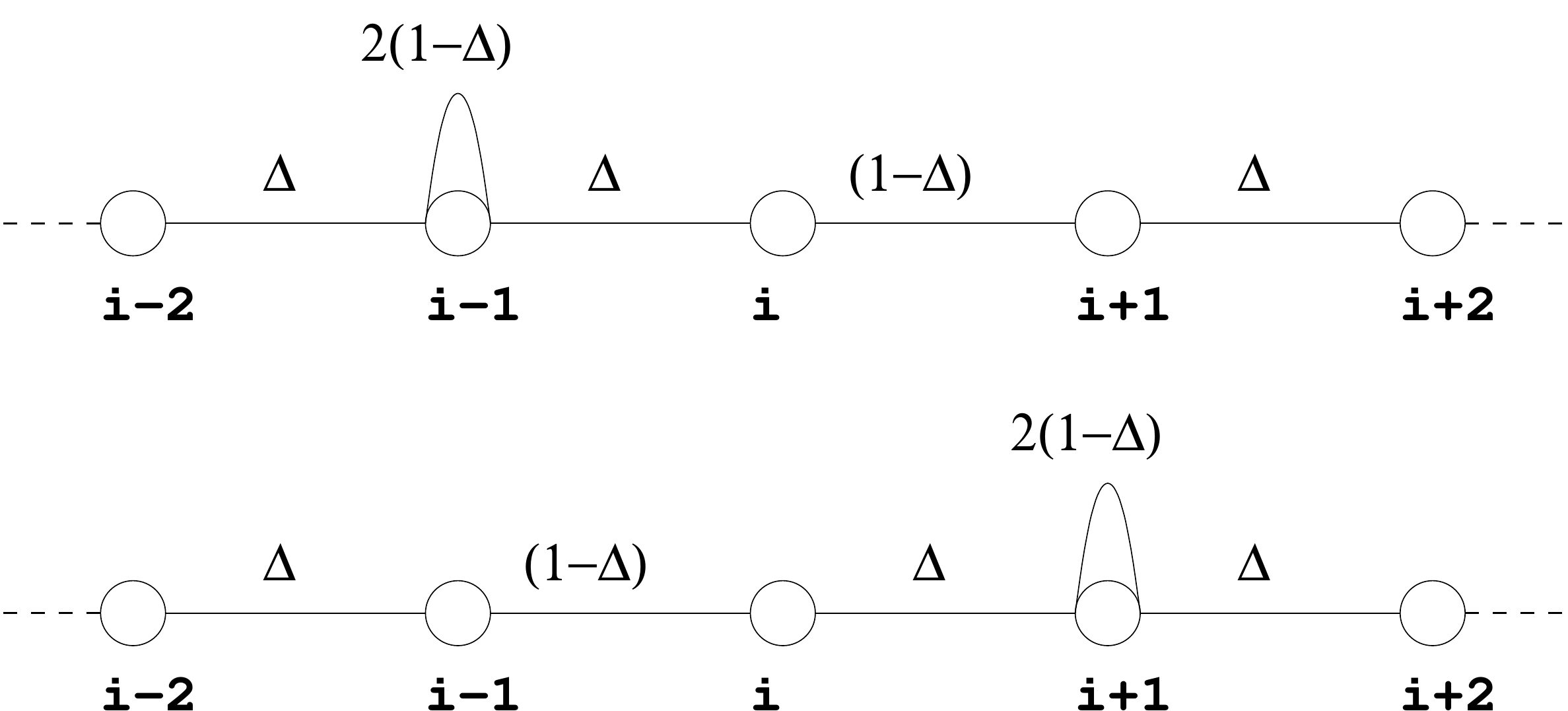}
\end{center}
\caption{Quiver diagram for the second type of the Seiberg duality}
\label{ciccio2}
\end{figure}

This result is proven by distinguishing two cases.
In the first case, shown in figure \ref{ciccio1}, the theory does not posses
adjoint fields in the first phase and after duality two adjoints arise.
In the second case there is one adjoint field and the duality acts as in figure \ref{ciccio2}.

Let us discuss the first case in more detail, 
where in the quiver before duality there are no adjoint fields (at least next to the 
group which undergoes the duality). The dual theory has instead two adjoint fields 
on the nodes $N_{i\pm1}$ if the $U(N_i)$ group is dualized.
The superpotential 
\begin{eqnarray}
W &=& X_{i-2,i-1}X_{i-1,i}X_{i,i-1}X_{i-1,i-2}-X_{i-1,i}X_{i,i+1}X_{i+1,i}X_{i,i-1} \nonumber \\ 
&+&X_{i,i+1} X_{i,i+2}X_{i+2,i+1}X_{i+1,i}+\Delta W
\end{eqnarray}
becomes
\begin{eqnarray}
W &=& X_{i-2,i-1} Y_{i-1,i-1} X_{i-1,i-2}-Y_{i,i-1}Y_{i-1,i-1}Y_{i-1,i}+Y_{i-1,i}Y_{i,i+1}Y_{i+1,i}Y_{i,i-1}
 \nonumber  \\&-&
Y_{i,i+1}Y_{i+1,i+1}Y_{i+1,i}+X_{i+2,i+1}Y_{i+1,i+1}X_{i+1,i} +\Delta W
\end{eqnarray}
where in $\Delta W$ we collected all the superpotential terms which are not involved in the duality.
Moreover there is a relation between the $R$ charges $\Delta$ in the two phases.
Indeed the adjoint fields $Y_{i\pm1,i\pm1}$ are related to the original fields as
\begin{equation}
Y_{i\pm 1,i\pm1} = X_{i\pm1,i}X_{i,1\pm 1}
\end{equation}
and the R charges become
\begin{equation} \label{RQUIV}
\widetilde \Delta_{i\pm1,i\pm1} = \Delta_{i,i\pm1}+\Delta_{i\pm1,i} = 2  \Delta_{i,i\pm1}
\end{equation}
where the last equality follows from the symmetry of the $\widetilde{L^{aba}}$
quivers. 
From (\ref{RQUIV}) and from the constraints imposed by the superpotential the other $R$ charges are 
assigned as in the figure \ref{ciccio2}. The fields which are not directly involved in the duality (mesons and dual quarks) 
have the same $R$ charge in both the theories.
At this stage of the discussion one can  apply the rules (\ref{rulesy}) and check that even the matter content of the 
dual theories gives the same contribution to the free energy.
We distinguish three sectors: the fields charged under $N_i$, the adjoints and the bifundamentals uncharged under
$N_i$, and we show that each sector separately contributes with the same terms.

The two pairs  of bifundamental fields $X_{i,i\pm 1}$ and $X_{i\pm1,i}$ contribute to the free energy as
\begin{equation}\label{FF1}
\Delta F_i = -(1-\Delta) \int \rho^2 (\delta y_{i-1,i}^2 -\frac{4}{3} \pi^2 \Delta(2-\Delta))
-\Delta \int \rho^2 (\delta y_{i,i+1}^2 -\frac{4}{3} \pi^2 \Delta(1+\Delta))
\end{equation}
while in the dual phase this contribution is
\begin{equation}\label{FDD2}
\Delta F_i = -\Delta \int \widetilde \rho^2 ( \widetilde \delta y_{i-1,i}^2 -\frac{4}{3} \pi^2  \Delta(1+\Delta))
-(1-\Delta)\int \widetilde \rho^2 (\widetilde \delta y_{i,i+1}^2 -\frac{4}{3} \pi^2 \Delta(2-\Delta))
\end{equation} 
The rules (\ref{rulesy}) map the new $\tilde y$ variables in the former ones as
\begin{equation} \label{relation}
\widetilde \delta y_{i\pm1,i} = - \delta y_{i\mp1,i}
\end{equation}
By substituting (\ref{relation}) in (\ref{FDD2}) formula (\ref{FF1}) is recovered (with
$\rho=\tilde \rho$ as in $F_{cl}$).

The second contribution to the one loop free energy comes from the adjoint fields. In the electric theory 
this contribution vanishes, because there are no adjoints for the nodes $N_{i\pm1}$.
In the dual theory there are two adjoint fields and their contribution is
\begin{equation}
\Delta F_{i\pm1} = \frac{2}{3}\pi^2 \sum_{\alpha=1}^{2}
\Delta_\alpha (1-\Delta_\alpha )(2-\Delta_\alpha)  \int \rho^2 dx
\end{equation}
In this case $\Delta_{1} =2 \Delta$ and $\Delta_2=2(1-\Delta)$
and the sum is vanishing, as in the other phase.

The last contribution comes from the other matter fields. The integrals
are the same in both the phases and the relation (\ref{SD})
guarantees that $\delta y_{\alpha \beta}=\widetilde \delta y_{\alpha \beta} $
if $\alpha\neq i \neq\beta$.  This proves that the dual theories have the same 
$F$ even before the extremization.

The second case is similar to the former one and we refer to it in the figure \ref{ciccio2}.
By repeating the analysis on the superpotentials above one finds a distribution 
of $R$ charges as in figure \ref{ciccio2}.
Then the analysis in straightforward. 
Indeed the contributions of the CS term and the one loop contribution of the bifundamental fields
are exactly as before, while the contribution
from the adjoints is trivially the same, since there is no $y$ dependence for the adjoints
and they have the same $R$ charge.

\section{Chiral-Like models}
\label{sec:Chirals}

In the vector-like models 
we observed that 
the field theoretical quantities
and their gravity duals match,
corroborating the validity of the conjectured AdS/CFT duality
for these models.
In the chiral case the situation is different.
Indeed, it is natural to apply the same
techniques explained above to these cases, but both analytics and numerical computations
do not match with the volume computations. 
In particular, the ansatz (\ref{eq:eigen}) is no longer valid, and as a result the free energy does not
respect the supergravity dual prediction $F\propto N^{3/2}$, but $F \propto N^2$. This of course does not
necessarily imply that the conjectured duality is ruled out, because the large $N$ saddle point approximation relies on
the assumption that we have identified a global minimum of the free energy.
A solution whose corresponding free
energy is proportional to $N^{3/2}$ would of course be energetically favored. 
We indeed found such a favored scaling in many examples numerically. 
Here we report on two cases, where we could reproduce the extremized value of the free energy also analytically.

\subsection{Symmetrizing the free energy: vectorialization}

In this section we compute the large $N$ free energy of some chiral models
by  applying  the symmetrization technique introduced in \cite{Amariti:2011jp}. 
The quiver gauge theories that we consider are   $\widetilde{\mathbb{F}_0^{(I)}}_{\{k,k,-k-k\}}$  and 
$\widetilde{\mathbb{C}^3/\mathbb{Z}_3}_{\{k,k,-2k\}}$, whose toric diagrams correspond to the  
$Q^{222}/\mathbb{Z}_k$ and $M^{111}/\mathbb{Z}_k$ models respectively. From now on we 
will denote these models with the corresponding Sasaki-Einstein space.

In these examples we checked that it is completely equivalent to fully symmetrize the integrand of the partition function or to only make
explicit a $\mathbb{Z}_2$ subgroup of the full symmetry group. Moreover, because the models only contain
bifundamental fields, we treat their matter contributions in a unified way, and later we will specialize to each example.

Let us consider the partition function (\ref{eq:Zgen}) for chiral-like models with $|G|$ gauge groups.
Call $N_f^{(a,a+1)}$ the number of bifundamental fields between group $a$ and group $a+1$.\footnote{The generalization
to more complicated models is straightforward.} While the
vector and Chern-Simons contributions remain unchanged under the symmetrization, the matter part 
becomes (up to a factor which gives a subleading contribution to the free energy in the large $N$ limit)
\begin{equation}
f_{mat} = \sum_i \chi^{(i)}
\label{eq:fmat}
\end{equation}
where we defined
\begin{equation}
\begin{split}
L_{ab}^{\rho\eta} \equiv \sum_{i,j} N_f^{(ab)} \, \ell & \left( 1-\Delta_{ab} + i \frac{\rho \l_i^{(a)} - \eta \l_j^{(b)}}{2 \pi} \right) \\
\xi^{(1)} \equiv \sum_{a=1}^{|G|} L_{a,a+1}^{++} \qquad \qquad \xi^{(2)} & \equiv \sum_{a=1}^{|G|} L_{a,a+1}^{--} \qquad \qquad
\chi^{(i)} \equiv \exp{\left( \xi^{(i)} \right)}
\end{split}
\label{eq:definitions}
\end{equation}
with $\rho,\eta=\pm 1$ and $N_f^{(ab)}=2 \,\, ($respectively $3)$ for every $a$ and $b$ in the case of $Q^{222}/\mathbb{Z}_k$ 
(respectively $M^{111}/\mathbb{Z}_k$).
Notice also that we have $L_{ab}^{--}=L_{ba}^{++}$ if $\Delta_{ab}=\Delta_{ba}$ and $N_f^{(ab)}=N_f^{(ba)}$.
The matter contribution to the free energy reads
\begin{equation}
F_m = - \log\left( \sum_i \chi^{(i)} \right)
\end{equation}
and its contribution to the equations of motion is
\begin{equation}
\frac{\delta F_m}{\delta \l_k^{(a)}} = - \left( \frac{1}{\sum_i \chi^{(i)}} \right) \left( \sum_i \chi^{(i)} \frac{\delta \xi^{(i)}}{\delta \l_k^{(a)}} \right) = - \frac{1}{n_b} \sum_{greatest\,\, i} \frac{\delta \xi^{(i)}}{\delta \l_k^{(a)}}
\label{eq:EOMmatter}
\end{equation}
The last step requires some explanation. In the large $N$ limit, every $\xi$ is divergent.
Nevertheless, we expect some of the $\chi$'s to be much larger than
the others (see the discussion in \cite{Amariti:2011jp} below equation (3.22)). Then the sum restricts over the $i$'s
corresponding to the largest $\chi^{(i)}$'s.\footnote{In our examples we only have two $\chi^{(i)}$'s. However,
equation (\ref{eq:EOMmatter}) also holds when we fully symmetrize the matter contribution. In this case, as explained in \cite{Amariti:2011jp},
all the combinations of $\rho$ and $\eta$ are allowed.}
The latter will of course be
equal to each other, but their derivatives,
in general, are not. $n_b$ represents the number of such greatest $\chi$'s.

We now introduce the simplifying assumption $\chi^{(1)}=\chi^{(2)}$. In turn, this implies either a constraint on the $R$
charges to be determined after the solution is found or a constraint on the eigenvalue distribution. According to the
discussion in \cite{Amariti:2011jp}, assuming that the eigenvalue distribution is symmetric implies $\chi^{(1)}=\chi^{(2)}$.
We will further discuss the validity of this assumption in the conclusions.

Then we write (\ref{eq:EOMmatter}) as
\begin{equation} \label{eq:main}
\frac{\delta F_m}{\delta \l_k^{(a)}} = - \frac{1}{2} \frac{\delta}{\delta \l_k^{(a)}} \sum_{b=1}^{|G|}  \left( L_{b,b+1}^{++} + L_{b,b+1}^{--} \right) = - \frac{1}{2} \frac{\delta}{\delta \l_k^{(a)}} \sum_{b=1}^{|G|}  \left( L_{b,b+1}^{++} + L_{b+1,b}^{++} \right)
\end{equation}
where $L_{b+1,b}^{++}$ represents the contribution of $N_f^{(b+1,b)}=N_f^{(b,b+1)}$ (fictitious) chiral superfields in the $(\bar N_b,N_{b+1})$ representation
of the gauge group with $R$ charge $\Delta_{b+1,b}=\Delta_{b,b+1}$. The last step in \eqref{eq:main} is justified by the observation below equation \eqref{eq:definitions}.
We now see that the contribution of
$N_f^{(b,b+1)}$ chiral superfields in the $(N_b,\bar N_{b+1})$ representation to the free energy is the same as the contribution
coming from $N_f^{(b,b+1)}/2$ pairs of chiral superfields in the (anti)bifundamental representation of the gauge groups.
Thus, we may apply the rules introduced in section \ref{sec:Fvector} for vector-like field theories. Note that this "vectorialization"
of the field theory closely resembles the one found in the weak coupling case \cite{Amariti:2011jp}, where it was observed that at
two loop order the contribution coming from a field in a representation of the gauge group is the same of that coming
from a field in the conjugate representation, even at finite $N$. It would be interesting to check whether this is true at
higher orders in perturbation theory and in the subleading contribution at strong coupling.

We now turn to the analysis of two explicit models.

\subsection{$Q^{222}/\mathbb{Z}_k$}

We first consider a field theory with gauge group $G=\prod_{a=1}^4 U(N)_{k_a}$ and four corresponding gauge fields.
Each gauge field comes with a Chern-Simons term with level $k_a$ such that $\sum_a k_a=0$. There are two bifundamental
fields connecting the $a$-th and $(a+1$ mod 4)-th node. They are denoted $X_{a,a+1}^{i}$, $i=1,2$. The matter content is summarized
in the quiver diagram in figure \ref{fig:quivABJM}. In our conventions, an oriented arrow connecting node $a$ to node $b$
denotes one of the fields $X_{ab}^i$ which is in the fundamental of $U(N)_{k_a}$ and in the antifundamental of $U(N)_{k_b}$.

The superpotential is given by (we always omit the coupling constants)
\begin{equation}
W = \epsilon_{ij}\epsilon_{lk} X_{12}^{(i)} X_{23}^{(l)} X_{34}^{(j)} X_{41}^{(k)}
\end{equation}

When the Chern-Simons levels are chosen to be $k_a=\left( k,k,-k,-k \right)$, the model is conjectured to describe
the low energy theory of M$2$ branes probing the Calabi-Yau singularity which is a cone over $Q^{222}/\mathbb{Z}_k$.
In this case we may describe the dual geometry by means of its toric diagram, which is given by the (\ref{toricQ222}).

We write the equations of motion \eqref{eq:main} and the corresponding free energy functional according to the
rules outlined in section \ref{sec:Fvector}, with $\Delta_{ab}^+=2\Delta_{ab}$ and $\Delta_{ab}^-=0$. 
We do not report the details, because the free energy
resembles the $\widetilde{{\cal C}/\mathbb{Z}_2}$ case discussed in section \ref{sec:FvectorExamples},
with slightly different charge assignments.
We find that the extremal value of $F$, where all the fields have dimension $1/2$
matches with the volume of the  $\mathbb{Z}_k$ orbifold of $Q^{222}$
\begin{equation}
\text{Vol}(Y) = \frac{\pi^4}{16 k}
\end{equation}
We did not observe a matching of $F$ with the geometry \emph{before} extremization.
Note that even tough the results for the vector-like models seem to suggest the full equivalence,
no explanation like \cite{Butti:2005vn,Eager:2010yu} for this possibility has been given so far.
Furthermore, in the well understood vector-like theories, the off-shell eigenvalue distribution is not symmetric anymore.
Since the symmetry of the eigenvalues is the motivation for the simplifying assumption on the $\chi$'s, 
we are not too surprised to find the coincidence with the volumes only after extremalization.

\subsection{$M^{111}/\mathbb{Z}_k$}
\begin{figure}
\centering
\includegraphics[width=.3\textwidth]{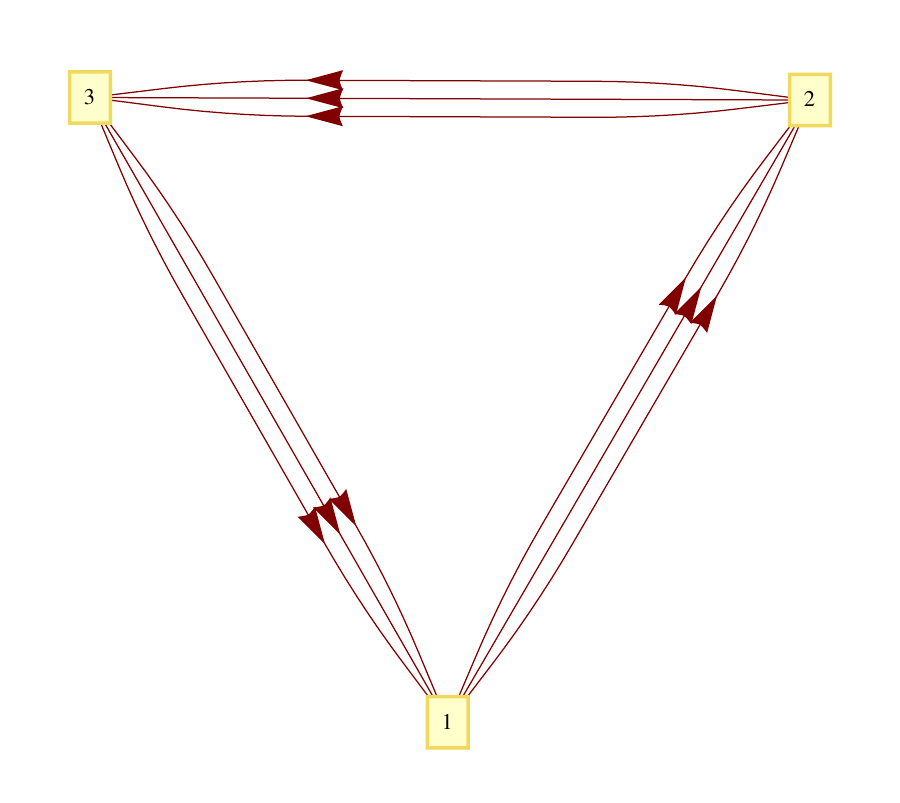}
\caption{Quiver diagram for $M^{111}$.}
\label{fig:M111}
\end{figure}
Our next example contains three gauge groups and three bifundamental fields connecting each pair of nodes, as
shown in Figure \ref{fig:M111}. The gauge fields still come with a Chern-Simons term and the levels sum up to zero.
The  superpotential reads
\begin{equation}
W = \epsilon_{ijk} X_{12}^{(i)} X_{23}^{(j)} X_{31}^{(k)}
\end{equation}
where the bifundamental fields $X_{a,a+1}$ are
in the fundamental of $U(N)_{k_a}$ and in the antifundamental of $U(N)_{k_{a+1}}$. 
If the CS levels are chosen to be $(k,k,-2k)$ the toric diagram is specified by (\ref{toricM111}).

In this case the free energy has no vector-like counterpart. We use the rules in section \ref{sec:Fvector}
with $\Delta_{ab}^+=2\Delta_{ab}$ and $\Delta_{ab}^-=0$ and determine the functional under
the assumption that the eigenvalue distribution is symmetric. On the extremal locus, where all 
fields have $R$ charge $2/3$, the volume of the dual Sasaki-Einstein manifold
is according to \eqref{eq:FVol} 
\be
{\rm Vol}(Y)= \frac{9 \pi^4}{128 k}
\ee
which agrees with the volume of a $\mathbb{Z}_k$ orbifold of $M^{111}$. 

\section{An alternative formula}
\label{sec:Rquartic}

In this section we look for a three dimensional  generalization of \cite{Butti:2005vn}, in which it was shown that 
the volume minimization is equivalent to the $a$-maximization in field theory.
Indeed in \cite{Butti:2005vn}
the authors computed the geometrical  $R$ 
charges from the toric data and provided a formula for the $a$-function in terms
of these  \emph{geometrical} $R$ charges and of the toric diagram.
Then in  \cite{Lee:2006ru} it was observed that this formula could be simplified
by imposing the constraints of the superpotential. The final result for the geometrical version
 of the $a$-function is
 \begin{equation} \label{formula21}
 a_g = \frac{1}{4} \sum \langle v_i,v_j,v_k\rangle R_i R_j R_k
 \end{equation}
 where $v_i$ represent the external points of the toric diagram, $\langle \cdot,\cdot,\cdot \rangle$ is the area of the oriented 
 surface generated by every set $i,j,k$ of external points and $R_i$ are the $R$-charges associated to each
 point of the diagram, which represent the set of fields in a given perfect matching.
This  cubic formula corresponds to the sum of the areas among the external points
weighted by the $R$ charges of their PM.

 One may be tempted to extend \eqref{formula21} to the three dimensional case. Here the 
 field theory candidate for the matching with the geometric data is the free energy. 
 Then, our candidate geometrical version of the free energy, $F_g$, is
 \begin{equation} \label{formula22}
F_g^2 = \frac{1}{6}\sum \langle v_i,v_j,v_k,v_l\rangle R_i R_j R_k R_l
 \end{equation}
 
 We observe  that \eqref{formula22} reproduces the 
 $Z$ function only if  there aren't internal  lines or surfaces 
in the toric diagram.  
Instead, if there are internal lines or surfaces, 
we didn't find any example in which  \eqref{formula22} reproduces the 
 $Z$ function.
Surprisingly, by adding some contribution related to the internal lines and surfaces 
 we have reproduced  the geometric $Z$ as a function of the Reeb vector.
 We  have not found a derivation for a general formula but we will  show the equivalence in several examples.
 The most interesting result is that  in all the examples only a quartic correction  
 in the $R$ charges associated to the external points of the toric diagram   is needed in order to identify $F_g^2\simeq1/Z$.

We therefore conjecture that this can always be done. If proven to be true, and
if the corrected $F_g$ equals the field theoretical free energy, our discussion would
offer a simpler extremization problem than the large $N$ limit of the localized free energy. However, two comments are in order.

It is important to stress that 
this  relation between $F_g^2$ and $Z$ does not involve any information about the dual field theory and applies 
directly to the toric diagram. 
This implies that it is not necessary to know the field theory dual 
but only the geometry of the mesonic moduli space to state the correspondence between $F_g^2$ and $Z$.
It  follows that the $F_g$ function that we define cannot solve the problems discussed in \cite{Jafferis:2011zi}
for the large $N$ scaling of the free energy in chiral theories.  

Another observation is that here  we simply define the $R$ charges of the perfect matchings associated to the external point
 of the toric diagram, and we do not relate them to any field theory description. With this procedure our candidate $F_g$ is always 
 polynomial in the $R$ charges  contrary with the known examples computed in the literature \cite{Martelli:2011qj,Jafferis:2011zi} 
and in section \ref{sec:Fvector},
where the free energy at large $N$ is a rational function of the $R$ charges of the fields and monopoles.
Anyway we checked in every example that the large $N$ free energy and our geometrical $F_g$
coincide once the symmetries among the perfect matchings are imposed.

\subsection{Examples}
The first case that we discuss is $D_3$.
In this case the toric diagram is
 \begin{equation} \label{toricSPP1-10}
 G_T=\left(
 \begin{array}{cccccc}
 0&1&0&1&0&1\\
 0&0&1&1&0&0\\
 1&1&0&0&0&0\\
 1&1&1&1&1&1\\
  \end{array}
  \right)
 \end{equation}
 The $Z$ function in terms of the Reeb vector is
 \begin{equation}
Z=\frac{ 16}{(b_1 b_2 b_3  (4 + b_1 - b_2)   ( b_3-4)  )}
 \end{equation} 
 The R charges associated to the six external points become $R_i=2 {\rm Vol}(\Sigma_i)/Z$.
 In this case we find that the conjectured geometrical free energy becomes
 \begin{equation} \label{geomF}
F_g^2=\frac{1}{6}  \sum \langle v_i,v_j,v_k,v_l\rangle R_i R_j R_k R_l=
 \frac{ b_1 b_2 b_3 (4 + b_1 - b_2)   ( b_3-4) }{16} 
  \end{equation}
and $F_g^2 Z=1$. 
Even if the (\ref{geomF}) is a polynomial function while (\ref{FD3final}) is a rational function they match once 
the symmetries among the PM are imposed.

Consider the general class of toric diagrams 
\footnote{Up to $SL(4,\mathbb{Z})$ transformations this class generalizes to every example of
the class $\mathbb{C}^2\times \mathbb{C}$ where the $\mathbb{C}^2 $ basis
refers to a four-dimensional parent theory with four external points.}
\begin{equation}
G_T=\left(
\begin{array}{ccccc}
x_1&0&0&x_4& 0\\
x_2&0&x_3&0&0\\
0&0&0&0&x_5\\
1&1&1&1&1
\end{array}
\right)
\end{equation}
with the constraint coming from the convexity 
$x_1x_3 + x_2 x_4-x_3 x_4>0$.
The $Z$ function is given by
\footnotesize
\begin{equation}
Z=\frac{((4 x_5 (x_2 ((b_2 - 4 x_5) x_1 x_3 + b_1 x_5 ( x_3-x_2 )) x_4 + 
    b_3 x_5 x_1 x_3 ( x_4-x_1 )))}{(
 b_1 b_2 b_3 (b_3 x_5 x_1 + (b_2 - 4 x_5) x_1 x_3 + 
    b_1 x_5 ( x_3-x_2 )) (b_3 x_5 (x_4-x_1 ) + x_2 (b_1 x_5 + (b_2 - 4 x_5) x_4))))}
\end{equation}
  \normalsize
By choosing $x_i>0$ and $x_1x_3 + x_2 x_4-x_3 x_4>0$ the geometric $F_g^2$ becomes
\footnotesize
 \begin{equation} 
F_g^2=\frac{1}{6}\frac
  {(
3 b_1 b_2 b_3 (b_3 x_5 x_1 + (b_2 - 4 x_5) x_1 x_3 + 
    b_1 x_5 ( x_3-x_2 )) (b_3 x_5 (x_4-x_1 ) + x_2 (b_1 x_5 + (b_2 - 4 x_5) x_4))))}
  {((2 x_5 (x_2 ((b_2 - 4 x_5) x_1 x_3 + b_1 x_5 ( x_3-x_2 )) x_4 + 
    b_3 x_5 x_1 x_3 ( x_4-x_1 )))}
  \end{equation}
  \normalsize
   and again $Z F_g^2=1$.
  
We now move to a vector-like example which requires a   correction.
   It can be obtained by modifying the toric diagram of the D$_3$ theory. 
   We consider a basis with four points $(0,0,0)$ $(1,0,0)$ $(0,1,0)$ and $(1,1,0)$
   as in D$_3$ but we modify the two points in the $z$ directions, such that 
   they are not associate to the splitting of two points on the same line
   on the plane $(x,y)$. This is not associated to an $SL$
   transformation and the toric diagram should describe a different model (for example 
   it can by obtained by an appropriate un-higgsing of the ABJM model). 
   The toric diagram is
  \begin{equation}
  G_T=\left(
  \begin{array}{cccccc}
  0 &0 &1& 1  &0& 1\\
  0 &1 &0 &1 &0 &1\\
  0 &0& 0 &0& 1& 1\\
    1 &1& 1 &1& 1& 1
    \end{array}
  \right)
  \end{equation}
The $Z$ function is 
  \begin{equation}
Z=\frac{  (16 ((b_1 - b_2)^2 - 16) + 
 8 (8 + b_1 b_2 - 2 b_1 - 2 b_2) b_3)}{(b_1 b_2 b_3 (b_1 - 4)  (b_2 - 4)  (b_1 - 
   b_2 - b_3 + 4)  (b_1 - b_2 + b_3 - 4))}
  \end{equation}
  while $F_g^2$, if computed from (\ref{formula22}), does not reproduce the expected result and it is a complicate 
  expression. 
  However we observe that   in this case  there exist two internal lines, connecting the points $1,6$ and $4,5$,
and an internal plane which passes through the points $1,4,5$ and $6$. As we discussed above
a correction proportional to 
  $\Delta F_g^2= -2 (R_1 R_6-R_4 R_5)^2$  can be added to $F_g^2$. With this correction it is straightforward to observe that 
  $F_g^2$ and $1/Z$ match.

We can  consider another class of vector-like models in which  (\ref{formula22})
does not coincide with $Z$. We refer to this class as  $\widetilde{SPP}_{(-m-1,m,1)}$
 (with $m>0$). The toric diagram is given by
   \begin{equation}
  G_T=\left(
  \begin{array}{cccccc}
  0 &2& 0& 1& 1& 1\\
  0& 0& 1& 1& 0& 0\\
  0& m-1& 0& 0& -1& m \\
    1& 1& 1& 1& 1& 1
    \end{array}
  \right)
  \end{equation}
In this case the $F_g^2$ function reproduces the $Z$
function only after the deformation 
\begin{equation} \label{corrSPP}
\Delta F_g^2 = 
\frac{4 
 (m R_1^2 R_4^2 - 2 R_1 R_2 R_3 R_4 + m R_2^2 R_3^2)}{m+1}
\end{equation}
  It is interesting to observe that the formula is still quartic and the deformation 
of $F_g$ involves all the sets of  coplanar and collinear external points .

The last examples that we  analyze are associated to the chiral-like cases
 investigated in the paper, 
 $M^{111}$ and $ Q^{222} $.
If the intuition that we got from the other examples 
is correct one must add a contribution proportional to all the possible internal planes and lines, 
by a quartic combination of their charges.

Let us turn to the first of the two examples, where the toric diagram is
\begin{equation}
G_T=\left(
\begin{array}{ccccc}
1&0&-1&0&0\\
0&1&-1&0&0\\
0&0&0&k_1&-k_2\\
1&1&1&1&1
\end{array}
\right)
\end{equation}
with $k_1,k_2>0$.
This diagram reduces to $M^{111}/\mathbb{Z}_k$ for $k_1=k_2$.
We found that the  geometrical $F_g^2$  and the $Z $ function may be identified if a correction 
\begin{equation}
\Delta F_g^2= -\frac{4 (k_1+k_2)^3}{9 k_1 k_2}R_4^2 R_5^2
\end{equation}
is added to $F_g^2$ (where $R_4$ and $R_5$ refer to the points with $k_1$ and $-k_2$ splitting.\\
In the second case, $Q^{222}$, the expression for $F_g$ reduces to the $Z$ function
only after adding the  correction
\begin{equation}
\Delta F_g^2 = 4(R_1 R_2 +R_3 R_4+R_5 R_6)^2
- 8 (R_1^2 R_2^2 +R_3^2 R_4^2+R_5^2 R_6^2)
\end{equation}
We see that even in this case it is possible to express the free energy
as a set of quartic combinations of the R charges.

It would be interesting to find a derivation of this result like in \cite{Butti:2005vn}  
and to see if it provides, at least in the toric case,  a different way for the computation of the free energy
instead of the localization of \cite{Jafferis:2010un}.

\section{Discussion and future directions}
\label{sec:discussion}

In this paper we have given further evidence for some conjectured AdS/CFT dual pairs. We matched the supergravity
computation with the field theoretical evaluation of the large $N$ free energy  of
vector  like toric quiver gauge theories. We also checked that the RG flows predicted by partial resolutions of the toric diagram
are in agreement with the conjectured $F$-theorem.

Then we studied the behavior of the free energy in the infinite family of $\widetilde L^{aba}_{k_i}$
models. We showed that at large $N$ the partition function is preserved among the Seiberg/toric dual phases,
where the rules of this duality where originally derived in \cite{Aharony:2008gk,Giveon:2008zn,Amariti:2009rb}. 

In the second part of the paper we focused on the free energy of chiral-like quiver gauge theories.
Even if these models are conjectured to be dual to M-theory on $AdS_4 \times X_7$,
the large $N$ scaling  of the free energy has not been observed \cite{Jafferis:2011zi}.
Here we  applied a recent proposal to evaluate in the saddle point approximation the free energy
of some chiral-like field theories \cite{Amariti:2011jp}. 
We observed that the expected scaling and the on-shell volume 
obtained from the supergravity computation can be recovered with our method in the cases $Q^{222}/\mathbb{Z}_k$ and
$M^{111}/\mathbb{Z}_k$.

In the last part of the paper we commented on a different quantity that can 
give the information on the exact $R$ charge in field theory. The construction is
based on the four dimensional relation among the $Z$ function and the $a$ function.
We constructed a field theoretical 
quantity which, in many examples, matches with the volume computation even before extremization.

We leave many open problems and we hope to come back on them in future publications.
First, one can extend the relation among the free energy and the 
Hilbert series even to non toric theories, as
observed in \cite{Eager:2010yu} for the $a$-maximization of \cite{Intriligator:2003jj}.

Another  extension of our work  is the role of the subleading contributions in the dualities that 
we checked here. Indeed, as we observed, the finite $k$ contribution in the dual gauge group
gives a leading contribution at large $N$ which cancels because the theory is vector like. A deeper check 
should consist of matching  the subleading contributions between the dual phases.
Moreover one should study the existence of similar dualities among theories
completely unrelated in four dimensions. Usually the dual phases are obtained by unhiggsing.
In many cases the unhiggsing involves a bifundamental field and a chiral like theory is generated,. 
Anyway by unhiggsing an adjoint field the daughter theory is still vector like. For example this is the case 
for the third phase of D$_3$ discussed in \cite{Davey:2009sr}.
This duality relates the classical mesonic moduli spaces, but a better check should be the matching 
of the free energy at large $N$.

An interesting result of the paper is the computation of the free energy at large $N$ in the chiral like theories.
Anyway in this case we left many open problems that deserve further studies.
Indeed the symmetrization approach yet suffers on a limitation. In all known examples, the eigenvalue distribution
has been shown to be a symmetric function  after extremization with respect to the $R$ charges. 
As we saw, when one assumes this, the terms contributing
to the matter part of the free energy are equal to each other. In turn, this implies that the contributions
from the monopole charges cancel out. 
Thus, with this assumption we can trust our free energy computation only for theories with expected vanishing monopole charge,
which is why we had to restrict our analysis to few models.
Relaxing the above assumption, and considering more
general models, is a hard task, but it is necessary to further check the relation with the geometry and to check the conjectured AdS/CFT duality
for these cases.

A further model with no expected monopole charge is the field theory dual to $AdS_4 \times Q^{111}/\mathbb{Z}_k$.
It is easy to see that the symmetrized free energy for this model satisfies the same equations as the generalized
conifold with levels $(k,0,-k,0)$. Thus, we find a solution such that the free energy shows the expected
scaling, namely $F \propto N^{3/2}$, but such that the field theory computation
does not match with the supergravity one. It remains an open problem whether this argument rules out
the conjectured duality or whether it is a shortcoming of the applied saddle point technique.
Finally, we would like to comment on the dual phases of our models proposed in \cite{Amariti:2009rb,Franco:2009sp,Davey:2009sr}.
Strictly speaking, these dualities have been derived
for vector-like models, but they were also shown to be applicable to some chiral-like theories, namely $Q^{111}/\mathbb{Z}_k$
and $Q^{222}/\mathbb{Z}_k$. The latter has a toric phase which is the analog of the Phase II of $F_0$ in four dimensions. We
applied our procedure to this phase as well, and got a result which differs from the  expected  one shown in section \ref{sec:Chirals}.
We hope to clarify this mismatch in future works.

Notice that the full understanding of the quantum corrected moduli space of chiral theories is intricate.
It certainly would be rewarding to see if an extension of the field theory models along the lines of \cite{Benini:2011cma} shed
light on some of the open problems reported here.

Another observation regards the geometrical version of the free energy proposed in section \ref{sec:Rquartic}.
We observed that, when the three dimensional toric diagram has internal lines or surfaces, 
some corrections must be added such that  $F_g^2$ is the inverse of  $Z$.
Anyway we do not have either a general procedure or a derivation of the formula, and we only found out the corrections in every examples by hand. It would be important to 
find a deeper origin for our claim and shed new light on the relation among $a$ and $F$ maximization.
Another subtle point which requires more investigation is the relation among the free energy computed in field theory and 
$F_g$ itself. Indeed as we already observed the first one is usually a rational function of the $R$ charges of the PM while the second one is 
by construction a polynomial function of the $R$ charges. In all the examples we studied they coincide up to the symmetries among the $\Sigma_i$ cycles wrapped by the M$5$ branes.
Anyway it is still unclear if there exists a purely polynomial version of the (large $N$) free energy  as a function of the $R$ charges in every 
$\mathcal{N}=2$ three dimensional field theory.

\section*{Acknowledgments}

It is a great pleasure to thank Alberto Zaffaroni for 
collaboration during the early stage of this project and
for many crucial hints and discussions. 
We are  also grateful to Ofer Aharony, Cyril Closset and Kenneth Intriligator for comments.
M.S. thanks the Universit\`a di Milano-Bicocca for hospitality.
A.A. is supported by UCSD grant DOE-FG03-97ER40546.  
The work of C.K.  is supported in part by INFN and in part by MIUR.
The work of M.S. is supported in part by the FWO - Vlaanderen,
Project No. G.0651.11, in part by the Federal Office for Scientific,
Technical and Cultural Affairs through the ``Interuniversity Attraction
Poles Programme -- Belgian Science Policy'' P6/11-P, and in part by the European Science Foundation Holograv
Network.

\appendix

\section{The $Z$-function for arbitrary Reeb vector}
\label{sec:Zfunction}

Here we compute the volumes $12/\pi^4 Z(b_i)$ of the Sasaki-Einstein manifolds 
dual to the field-theoretical models we will be interested in the rest of the paper. In the case of toric manifolds, 
the computations only need the knowledge of the toric diagram and the volumes are rational functions of the
Reeb vector $\mathbf{b}=(b_i)_{i=1\dots4}$.
By identifying  $a_i(\mathbf{b}) =  2\mathrm{Vol}_{\Sigma_i}/Z$,
these results are in agreement with the ones discussed in paper.

\paragraph{$\mathbb{C}\times \mathcal{C}$.}
The toric diagram is shown in figure \ref{fig:CxConif}, it is spanned by the vectors 
\begin{equation}
G_t = 
\left(
\begin{array}{ccccc}
2&0& 0& 1& 1\\
0 &0&-1& 0& 0\\
1&0&0&1&0\\
1&1&1&1&1
\end{array}
\right) \, ,
\label{eq:ABtoric}
\end{equation}
and, applying the standard techniques discussed above, we find the volumes
at arbitrary Reeb vector
\begin{align}
\mathrm{Vol}_{\Sigma_1} &=   \frac{-1}{b_2 \left(4+b_2-b_3\right) \left(4-b_1+b_2+b_3\right)}\, ,\nonumber\\
\mathrm{Vol}_{\Sigma_2} &=   \frac{1}{b_2 b_3 \left(b_3-b_1\right)}\, ,\nonumber \\
\mathrm{Vol}_{\Sigma_3} &=   \frac{4+b_2}{\left(4+b_2-b_3\right) b_3 \left(b_1-b_3\right) \left(4-b_1+b_2+b_3\right)} \nonumber \, ,\\
\mathrm{Vol}_{\Sigma_4} &=  \frac{1}{b_2 \left(4+b_2-b_3\right) \left(b_3-b_1\right)} \nonumber\,, \\
\mathrm{Vol}_{\Sigma_5} &=   \frac{-1}{b_2 b_3 \left(4-b_1+b_2+b_3\right)}\, ,\nonumber\\
Z &= \frac{4 \left(4+b_2\right)}{b_2 \left(4+b_2-b_3\right) b_3 \left(-b_1+b_3\right) \left(4-b_1+b_2+b_3\right)} \, ,
\label{eq:VoloffCC}
\end{align}
with $Z  \equiv \sum_{i=1}^{5} \mathrm{Vol}_{\Sigma_i} = 12 \mathrm{Vol}(H) /\pi^4 $.

\paragraph{$\widetilde{SPP}$.}
The toric diagram
for the SPP is 
\begin{equation}
G_t = 
\left(
\begin{array}{cccccc}
0&0& -1& -1& 0&0\\
0 &2&0& 1& 1&1\\
0&k_2-k_3&0&0&-k_3&k_2\\
1&1&1& 1& 1&1
\end{array}
\right) \, ,
\label{eq:ABCtoric}
\end{equation}
shown in figure \ref{fig:SPP} for the case $k_2=k_3=-1$. 
Note that we have chosen a different $SL(4,\mathbb{Z})$ frame then in section \ref{sec:Rquartic}. We can compute the volumes
\begin{align}
\label{eq:ABCvols}
\mathrm{Vol}_{\Sigma_1} &= \frac{k_2+k_3}{b_1 \left(b_2 k_2-b_3\right) \left(b_3+b_2 k_3\right)} \, ,\nonumber\\
\mathrm{Vol}_{\Sigma_2} &= \frac{k_2+k_3}{b_1 \left(b_3+\left(4-b_2\right) k_2+\left(4+b_1\right) k_3\right) \left(\left(4+b_1\right) k_2+\left(4-b_2\right) k_3-b_3\right)} \, ,\nonumber \\
\mathrm{Vol}_{\Sigma_3} &= \frac{\left(k_2+k_3\right) \left(b_3 \left(k_2-k_3\right)+\left(4+b_1+b_2\right) k_2 k_3\right)}{\left(b_3-\left(4+b_1\right) k_2\right) \left(b_2 k_2-b_3\right) \left(b_3+\left(4+b_1\right) k_3\right) \left(b_3+b_2 k_3\right)} \nonumber \, ,\\
\mathrm{Vol}_{\Sigma_4} &= \nonumber
 \textstyle  \frac{-\left(k_2+k_3\right) \left(\left(4+b_1\right) k_2^2+\left(4-b_2\right) k_2 k_3+\left(4+b_1\right) k_3^2+b_3 \left(k_3-k_2\right)\right)}{\left( b_3-\left(4+b_1\right) k_2\right) \left(b_3+\left(4+b_1\right) k_3\right) \left(b_3+\left(4-b_2\right) k_2+\left(4+b_1\right) k_3\right) \left(b_3-\left(4+b_1\right) k_2-\left(4-b_2\right) k_3\right)}  \nonumber\,, \\
\mathrm{Vol}_{\Sigma_5} &= \frac{b_3 \left(k_2+k_3\right)+k_3 \left(4 k_2+\left(4+b_1\right) k_3\right)}{b_1 \left(b_3+\left(4+b_1\right) k_3\right) \left(b_3+\left(4-b_2\right) k_2+\left(4+b_1\right) k_3\right) \left(b_3+b_2 k_3\right)} \, ,\nonumber\\
\mathrm{Vol}_{\Sigma_6} &= \frac{b_3 \left(k_2+k_3\right)-k_2 \left(\left(4+b_1\right) k_2+4 k_3\right)}{b_1 \left(\left(4+b_1\right) k_2-b_3\right) \left(b_3-b_2 k_2\right) \left(\left(4+b_1\right) k_2+\left(4-b_2\right) k_3-b_3\right)} \, .\nonumber
\end{align}
The formula for $Z$ is lengthy, we refrain from an explicit expression here.
Note, nevertheless the two special cases $D_3$, corresponding to CS levels $(1,-1,0)$,
and $SPP_{2-1-1}$,
\begin{align}
  Z_{D_3} &=  \frac{16}{b_1b_3\left(4+b_1+b_3\right)\left(4-b_2-b_3\right) \left(b_2+b_3\right)} \, , \nonumber \\
  Z_{SPP_{2-1-1}}  &= 
  \frac{8 \left(b_1^3+b_1^2 \left(20-b_2\right)+b_1 \left(128-8 b_2+b_2^2-3 b_3^2\right)+16 \left(16-b_3^2\right)\right)}
  {b_1 \left(\left(4+b_1\right)^2-b_3{}^2\right) \left(\left(8+b_1-b_2\right)^2-b_3^2\right)\left(b_3^2-b_2^2\right)}\, .
\end{align}

\paragraph{$\widetilde{{\cal C}/\mathbb{Z}_2}$.}
The toric diagram is spanned by
\begin{equation}
G_t = 
\left(
\begin{array}{cccccccc}
1&1& 0& 0& 0&0&1&1\\
0 &-2&0& -2&-1&-1&-1&-1\\
0&0&0&0&-1&1&0&0\\
1&1&1&1&1&1&1&1
\end{array}
\right) \, ,
\label{eq:ABCDtoric}
\end{equation}
which gives the volumes
\begin{align}
\label{eq:ABCDvols}
\mathrm{Vol}(\Sigma_1) &=   \frac{2 \left(4-b_1-b_2\right)}{\left(\left(4-b_1\right){}^2-b_3^2\right) \left(b_2^2-b_3^2\right)} \, ,\nonumber\\
\mathrm{Vol}(\Sigma_2) &=   \frac{2 \left(12-b_1+b_2\right)}{\left(\left(4-b_1\right){}^2-b_3^2\right) \left(\left(8+b_2\right){}^2-b_3^2\right)} \, ,\nonumber \\
\mathrm{Vol}(\Sigma_3) &=   \frac{2}{b_1 \left(b_2^2-b_3^2\right)} \nonumber \, ,\\
\mathrm{Vol}(\Sigma_4) &=   \frac{2}{b_1 \left(\left(8+b_2\right){}^2-b_3^2\right)} \nonumber\,, \\
\mathrm{Vol}(\Sigma_5) &=  \frac{2 \left(4+b_3\right)}{b_1 \left(4-b_1+b_3\right) \left(b_3-b_2\right) \left(8+b_2+b_3\right)} \, ,\nonumber\\
\mathrm{Vol}(\Sigma_6) &=   \frac{-2 \left(4-b_3\right)}{b_1 \left(4-b_1-b_3\right) \left(8+b_2-b_3\right) \left(b_2+b_3\right)} \,,\nonumber\\
Z &= \frac{18 \left(b_1 \left(32+8 b_2+b_2^2-b_3^2\right)+8 \left(b_3^2-16\right)\right)}
  { k \pi ^2b_1 ((4-b_1)^2-b_3^2) ((8+b_2)^2-b_3^2) (b_3^2-b_2^2) } \,.
\end{align}

\paragraph{ABJM$/\mathbb{Z}_2$.}
The toric diagram in figure \ref{fig:quivABJM} is spanned by
\begin{equation}
G_t = 
\left(
\begin{array}{cccccccc}
1&-1& 0 & 0& 0&0 &0 &0\\
0 &0 &1& -1&0&0&0&0 \\
2&0&0&0&1&1&0&0\\
1&1&1&1&1&1&1&1
\end{array}
\right) \, ,
\label{eq:ABJMtoric}
\end{equation}
which gives the volumes
\begin{align}
\label{eq:ABJMvols}
\mathrm{Vol}(\Sigma_1) &=   \frac{2}{\left(b_3-2 b_1\right) \left(4+b_1-b_2-b_3\right) \left(4+b_1+b_2-b_3\right)}\, ,\nonumber\\
\mathrm{Vol}(\Sigma_2) &=   \frac{2}{\left(4+b_1-b_2-b_3\right) \left(4+b_1+b_2-b_3\right) b_3} \, ,\nonumber \\
\mathrm{Vol}(\Sigma_3) &=   \frac{2}{\left(b_3-2 b_1\right) \left(4+b_1-b_2-b_3\right) b_3} \nonumber \, ,\\
\mathrm{Vol}(\Sigma_4) &=   \frac{2}{\left(4+b_1+b_2-b_3\right) b_3 \left(b_3-2 b_1\right)} \nonumber\,, \\
Z &= \frac{16}{\left(4+b_1+b_2-b_3\right) b_3 \left(b_3-2 b_1\right) \left(4+b_1-b_2-b_3\right)} \,.
\end{align}

\paragraph{$Q^{222}/\mathbb{Z}_k$.}

The toric diagram for $k=1$ is given by
\begin{equation}\label{toricQ222}
G_t = 
\left(
\begin{array}{cccccccc}
1&-1& 0& 0& 0& 0& 0& 0 \\
0 &0& 1& -1& 0& 0& 0& 0 \\
0&0&0&0&1&-1&0&0\\
1&1&1&1&1&1&1&1
\end{array}
\right)
\end{equation}
The volume of $Q^{222}/\mathbb{Z}_k$ is proportional to the minimum of
\begin{equation}
\label{ZQ222}
Z=\frac{32 ( b_1^2 + b_2^2 + b_3^2) + b_1^4 + b_2^4 + b_3^4 - 
 2 ( b_1^2 b_2^2 + b_1^2 b_3^2 + b_2^2 b_3^2) - 768}
{-\sqrt{ {\prod} _{\alpha,\beta,\gamma,\delta =\pm  1} 
\left(4\alpha  + \beta  b_1+ \gamma  b_1+ \delta  b_3\right)}}
\end{equation}
In this case no computation is actually needed: by the symmetry of the $Z$ function, the minimum is found for
$b_1=b_2=b_3 \equiv b$ and the variational problem further sets $b=0$. Then, the volume of the compact manifold
is given by
\begin{equation}
{\rm vol}(Q^{222}) =\frac{\pi^4}{12} Z = \frac{\pi^4}{16}
\label{eq:volQ222}
\end{equation}

\paragraph{$M^{111}/\mathbb{Z}_k$.}

The toric diagram for this geometry is specified by 
\begin{equation} \label{toricM111}
G_t = 
\left(
\begin{array}{cccccc}
1&-1&0&0&0&0\\
0&-1&1&0&0&0\\
0&0&0&1&-1&0\\
1&1&&1&1&1
\end{array}
\right)
\end{equation}
and the corresponding $Z$ function is
\begin{equation} \label{Zeta}
Z =- \frac{72 ^3 \left(b_3^2-3 \left(b_1^2-b_1 b_2+b_2^2-16\right) \right)}{\left(b_3^2-(4+b_1-2 b_2)^2 \right) \left(b_3^2-(4-2 b_1+b_2)^2 \right) \left(b_3^2-(4+b_1+b_2)^2 \right)}
\end{equation}
The $b_1$ and $b_2$ components of the Reeb vector $b=(4,b_1,b_2,b_3)$ can be fixed 
by the symmetries to be equal. By minimizing this function the components of 
$b$ are $b_1=b_2=b_3=0$ and all the $R$ charges of the fields become $2/3$ while the monopole
charge vanishes. The volume of the compact manifold $M^{111}$ is
\begin{equation}
{\rm vol}(M^{111} )=\frac{\pi^4}{12} Z = \frac{9 \pi^4}{128}
\end{equation}

\bibliographystyle{JHEP}
\bibliography{BibFile}

\end{document}